\documentclass[fleqn,usenatbib]{mnras}

\usepackage{newtxtext,newtxmath}

\usepackage[T1]{fontenc}
\usepackage{ae,aecompl}
\usepackage{bm}

\usepackage{graphicx}	
\usepackage{amsmath}	
\usepackage{amssymb}	

\newcommand{\st}{\mathrm{St}}

\title[Migrating Planets in Inviscid Dusty PPDs]{Migrating Low-Mass Planets in Inviscid Dusty Protoplanetary Discs}

\author[Hsieh \& Lin]{
He-Feng Hsieh,$^{1,2}$\thanks{E-mail: hfhsieh@gapp.nthu.edu.tw}
Min-Kai Lin$^{2}$
\\
$^{1}$Institute of Astronomy, National Tsing Hua University, Hsinchu 30013, Taiwan\\
$^{2}$Institute of Astronomy and Astrophysics, Academia Sinica, Taipei 10617, Taiwan\\
}

\date{Accepted XXX. Received YYY; in original form ZZZ}

\pubyear{2020}

\begin{document}
\label{firstpage}
\pagerange{\pageref{firstpage}--\pageref{lastpage}}
\maketitle

\begin{abstract}
Disc-driven planet migration is integral to the formation of planetary systems. In standard, gas-dominated protoplanetary discs, low-mass planets or planetary cores undergo rapid inwards migration and are lost to the central star. However, several recent studies indicate that the solid component in protoplanetary discs can have a significant dynamical effect on disc-planet interaction, especially when the solid-to-gas mass ratio approaches unity or larger and the dust-on-gas drag forces become significant. As there are several ways to raise the solid abundance in protoplanetary discs, for example through disc winds and dust-trapping in pressure bumps, it is important to understand how planets migrate through a dusty environment. To this end, we study planet migration in dust-rich discs via a systematic set of high-resolution, two-dimensional numerical simulations. We show that the inwards migration of low-mass planets can be slowed down by dusty dynamical corotation torques. We also identify a new regime of stochastic migration applicable to discs with dust-to-gas mass ratios $\gtrsim 0.3$ and particle Stokes numbers $\gtrsim 0.03$. In these cases, disc-planet interaction leads to the continuous development of small-scale, intense dust vortices that scatter the planet, which can potentially halt or even reverse the inwards planet migration. We briefly discuss the observational implications of our results and highlight directions for future work. 
\end{abstract}




\section{Introduction}

Observations of protoplanetary discs (PPDs) often reveal prominent dust gaps and rings \citep{andrews18,long18}. As dust grains and pebbles 
are intrinsic components of PPDs \citep{chiang10,testi14} and provide the raw material for planet formation \citep{johansen14,raymond20}, 
one naturally expects newly born planets to interact with solids in the disc \citep{johansen17}. Indeed, a promising explanation for the observed sub-structures in PPDs is disc-planet interaction  \citep{dong15,dipierro18,zhang18}. In this scenario, a planet induces a gas gap  \citep{lin93} and solids are trapped in pressure bumps at the gap edges  \citep{paardekooper04,paardekooper06b,fouchet07,ayliffe12,zhu12,dipierro17,Dong2017}. However, such models for observed dust rings often assume planets on fixed, circular orbits. 

On the other hand, disc-planet interaction also leads to orbital migration, 
which is pivotal in shaping the architecture of planetary  systems \citep{baruteau14,baruteau16,nelson18}. Planet migration has been studied for decades \citep{goldreich80} and contemporary models can include sophisticated physical effects such as magnetic fields   \citep{terquem03,guilet13,comins16}, turbulence,\citep{nelson04, uribe11,baruteau11c}, vortices \citep{li09,lin10,mcnally19} 
large-scale laminar flows \citep{mcnally17,kimmig20}, non-isothermal effects \citep{paardekooper06,paardekooper08,Paardekooper2010}, self-gravity \citep{baruteau08,zhang08}, etc. However, most studies of planet migration consider purely gaseous discs. 

The orbital migration of planets in dusty discs has been considered more recently \citep{mcnally19,Meru2019,nazari19,Perez2019,Weber2019,wafflard20}. These  studies focus on the the effect of a migrating planet on the morphology of a dusty PPD and find that a moving planet can induce multiple dust rings.  Furthermore, the migrating planet need not reside inside a dust gap at any given time. These new results have key observational implications. For example, multiple dust rings observed in PPDs need not correspond to multiple planets. However, a limitation of these models is that the dust particles are often treated as passive and do not actively affect planet migration. 

The effect of dust on planet migration was first considered by \cite{llambay18}. They found that,  
even for a canonical dust-to-gas mass ratio of $0.01$ \citep{chiang10},  
the dust can exert a positive torque on a planet that may halt inwards migration, which would otherwise pose a threat to the survival of planetary systems. Dust-related torques are expected to become more important with increasing solid abundance.  

There are several routes to enhance the solid-to-gas mass ratio in PPDs either globally or locally. These include magnetized disc winds \citep{bai17,wang19,gressel20}; photoevaporation \citep{alexander07,alexander14,wang17}; dust-trapping by pressure bumps \citep{pinilla12,pinilla17,dullemond18}; and mutual relative radial drift between dust and gas \citep{gonzalez17,kanagawa17}. It is therefore important to understand how planets interact with dust-rich discs \citep[e.g.][]{morbi20}. In particular, as the dust-to-gas mass ratio approaches unity, the back-reaction from dust drag onto the gas should be included.  

In a previous study, we simulated low-mass planets embedded in dust-rich discs and found that such planets experience large-amplitude, oscillatory torques related to the dust-gas relative drift in the planet's co-orbital region \citep{Chen2018}. Moreover, for sufficiently large dust grains and abundances, small-scale, dust vortices develop near the planet, which introduces additional torque variability. Such planet-induced, dust vortices have also been identified in other simulations \citep{Pierens2019,Yang2020} and may be related to the instability of thin dust rings \citep{Huang2020}. However, neither \citeauthor{Chen2018} nor \citeauthor{Yang2020} considered planet migration. \citeauthor{Pierens2019} ran a few simulations with a freely migrating planet and found in some cases vortices can reverse the planet's inwards migration and push it outwards. 

In this work, we examine in more detail the orbital migration of low-mass planets in dust-rich discs using high-resolution numerical simulations. We explore a range of dust abundances and degree of dust-gas coupling. We show that planet migration transitions from being smooth, steady, and inwards for tight coupling and/or low dust abundances; to being stochastic in the less-coupled and high dust abundance regime. The latter arises from the development of numerous dust vortices that continuously scatter the planet. Our simulations reaffirm the possibility of stopping the inwards migration and saving planetary cores in dusty discs. 

This paper is organised as follows. We first describe the disc-planet system of interest and our numerical model in \S\ref{model}. Our simulation results are presented in \S\ref{results}, where we describe the disc morphology and orbital migration as a function of dust parameters. We also briefly explore the the effect of viscosity, resolution, surface density profile, and planet mass. We discuss the observational implications of our results in \S\ref{discussions} along with simulation caveats to be improved in future work. We summarise and conclude in \S\ref{summary}. In the Appendices we present supplementary simulations to compare with \cite{Chen2018} and runs with prescribed planet migration rates to better understand some of our main results.

\section{Methods}\label{model}

We study the orbital migration of a low-mass planet in an isothermal, inviscid, and non-self-gravitating protoplanetary disc. We consider a two-fluid disc, consisting of gas and one dust species, in the two-dimensional (2D), razor-thin disc approximation, where the densities and pressure are replaced by the vertically integrated quantities. Both the gas and dust species are treated as fluids with surface density $\Sigma_\text{g}$ and $\Sigma_\text{d}$, respectively; but the dust fluid has zero pressure. The fluid approximation for dust is valid for small grains with $\st\ll 1$ \citep{jacquet11}, where $\st$ is the particle Stokes number defined below. Here, we describe the basic equations, fiducial disc models, and define torque formulae for later analyses. 

For convenience, we define a reference radius $r_0$ from the central star and use the subscript `$0$' to evaluation at $r_0$. We also use a subscript `p' to denote evaluation at the planet's position, which can be time-dependent. We take $r_0$ to be the initial orbital radius of the planet. 

\subsection{Governing Equations} 
We numerically solve the continuity and Navier-Stokes equations in the cylindrical polar coordinate, $(r, \phi)$, in a frame of reference centred on the central star:
\begin{align}
    &\frac{\partial \Sigma_\text{g}}{\partial t} + \nabla \cdot (\Sigma_\text{g} {\bf{u}}) = 0,  \\
    &\frac{\partial {\bf{u}}}{\partial t} + ( {\bf{u}} \cdot \nabla ) {\bf{u}}
  = - \nabla \Phi - \frac{\nabla P}{\Sigma_\text{g}} - \frac{\Sigma_\text{d}}{\Sigma_\text{g}} {\bf{f}}_\text{drag}, \label{gas_mom} \\
    &\frac{\partial \Sigma_\text{d}}{\partial t} + \nabla \cdot (\Sigma_\text{d} {\bf{v}}) = 0,  \\
    &\frac{\partial {\bf{v}}}{\partial t} + ( {\bf{v}} \cdot \nabla ) {\bf{v}} 
  = - \nabla \Phi + {\bf{f}}_\text{drag},
\end{align}
where $\bf{u}$ and $\bf{v}$ are respectively the gas and dust velocity,  $\Phi$ is the gravitational potential, and $\bf{f}_\text{drag}$ is the aerodynamic drag force between dust and gas. Note that the back-reaction, or feedback, from dust onto the gas is included in equation~(\ref{gas_mom}). 

We consider (locally) isothermal discs where the vertically integrated pressure $P = c_\text{s}^2\Sigma_\text{g}$. The sound-speed $c_\text{s} \equiv H\Omega_\text{K}$ is prescribed and stationary, where $H$ is the pressure scale height and $\Omega_\text{K} = \sqrt{GM_\star/r^3}$ is the Keplerian frequency, with $G$ and $M_\star$ being the gravitational constant and stellar mass, respectively. We consider a flared disc in which the aspect ratio is parameterised as $H/r \equiv h = h_0 (r / r_0)^f$, where $f$ is the flaring index.

The gravitational potential includes the direct potentials from the star and the planet, and the indirect term associated with the planet due to the non-inertial frame of reference:
\begin{equation}
    \Phi(\bm{r}) = -\frac{G M_\star}{r} 
         - \frac{G M_\text{p}}{ \sqrt{ \left| {\bf{r}} - {\bf{r}}_\text{p} \right|^2 + r_\text{s}^2 } }
         + \frac{G M_\text{p}}{r_\text{p}^2} r \cos \phi,
\end{equation}
where $M_\text{p}$ is the planet mass, ${\bf{r}}_\text{p}(t)$ is the relative position of the planet to the star, and $r_\text{p} = \left|\bm{r}_\text{p}\right|$. The softening length for the planetary potential is parameterized as $r_\text{s} = b H_\text{p}$, where $H_\text{p} = H( r_\text{p} )$ and $b$ is the smoothing factor. The disc's self-gravity and indirect potential are neglected since we consider low-mass discs. 

We assume the dust-gas drag force is in the Epstein regime in which its strength depends linearly on the relative velocity between the gas and dust \citep[see e.g.][]{Whipple1972}:
\begin{equation}
	{\bf{f}}_\text{drag} = \frac{\Omega_\text{K}}{\st} ({\bf{u}} - {\bf{v}}). 
\end{equation}
The Stokes number, $\st$, is a dimensionless measure of dust-gas coupling. Particles with $\st \ll 1$, as  considered in this work, are well-coupled to the gas. Physically, $\st$ depends on the properties of the dust species and surrounding gas as $\st =  \left(\pi/2\right)\left(a_\text{d}\rho_\text{d}/\Sigma_\text{g}\right)$, where $a_\text{d}$ and $\rho_\text{d}$ are the grain size and internal density, respectively \citep{weiden77}. However, for simplicity we consider a constant Stokes number for each simulation. In our simulations the variation in the gas surface density is at most of a factor of two from its initial values, so our results should not differ significantly from physically-prescribed Stokes numbers, e.g. based on fixed particle sizes.

The position and velocity of the embedded planet are updated by solving the equation of motion:
\begin{align}
%
	& \frac{d^2 {\bf r}_\text{p}}{d t^2} = - \frac{G (M_\star + M_\text{p})}{r_\text{p}^3} {\bf r}_\text{p} - \iint \frac{G\left(\bm{r}_\text{p} 
	- \bm{r}\right)\left(\Sigma_\text{g} + \Sigma_\text{d}\right)}{\left(\left|\bm{r}_\text{p} 
	- \bm{r}\right|^2 + r_\text{s}^2\right)^{3/2}}rdrd\phi,  
\end{align}
where the force contains the contributions from the star, the indirect term, and the dusty disc. The surface integral is taken over the entire disc without cut-offs near the planet. Unless otherwise stated, the planet is allowed to migrate from when it is introduced at $t=0$.

\subsection{Fiducial Model}\label{fid_model}

We initialize the 2D disc in an axisymmetric, steady state.
The initial gas surface density and squared sound-speed are given by
\begin{align}
	& \Sigma_\text{g} (r) = \Sigma_\text{g0} \left( \frac{r}{r_0} \right)^{-\sigma}, \\
	& c_\text{s}^2 (r) = c_\text{s0}^2 \left( \frac{r}{r_0} \right)^{-\beta},
\end{align}
where $\beta = 1 - 2 f$. The sound-speed is fixed in time. 
We follow \citet{Chen2018} and set the initial gas surface density and the squared sound-speed to be constant, $\sigma = \beta = 0$, which corresponds to a flaring index of $f = 0.5$. The radial pressure gradient of the gas disc thus vanishes everywhere so both gas and dust are in Keplerian motion, $v_\phi = u_\phi = r\Omega_\text{K}$. For inviscid gas, the radial velocity vanishes and there is no radial drift of dust \citep{weiden77}, i.e. $v_r = u_r = 0$. 

To set the gas surface density scale $\Sigma_\text{g0}$, we specify the vertically-integrated, radially-constant dust-to-gas ratio (or metallicity), $Z = \Sigma_\text{d} / \Sigma_\text{g}$, and Toomre parameter defined from the total surface density, i.e. 
\begin{equation}
    Q = \frac{c_\text{s} \Omega_\text{K}}{\pi G \Sigma_\text{g} (1 + Z)}.
\end{equation}
We adopt $Q_0 = 10$ and vary the metallicity in this study. 

To set the reference sound-speed $c_\text{s0}= h_0r_0\Omega_\text{K0}$, we recall that for well-coupled, small grains, dust-loading decreases the effective sound-speed of the dust-plus-gas system by factor of $\sqrt{1+Z}$ \citep{Laibe2014,Lin2017}. We can thus define an effective scale height 
\begin{equation}
    \widetilde{H} \equiv \frac{H}{\sqrt{1 + Z}}\label{Htilde}
\end{equation}
\citep{Chen2018} and an effective aspect ratio $\widetilde{h} = \widetilde{H}/r$. We choose the reference gas disc aspect ratio $h_0$ such that all simulations have $\widetilde{h}_0 = 0.05$ regardless of metallicity. 

We place one planet initially on a circular orbit with orbital distance $r_\text{p} = r_0$. The planet-to-star mass ratio is set to $q \equiv M_\text{p} / M_\star = 6 \times 10^{-6}$, which corresponds to a two Earth-mass planet orbiting a solar-mass star. In each simulation the planet's softening parameter $b$ is chosen such that its softening length $r_\text{s} = 0.6 \widetilde{H}_p $ initially.

Our adopted planet masses are below the feedback mass $M_\text{F}$, beyond which disc-feedback acts to slow down planet migration \citep[see][for its definition in a gas disc]{Rafikov2002b,mcnally19}. For a dusty disc where $h_\text{p}\to \widetilde{h}_\text{p}$ it is given by 
\begin{equation}
    M_\text{F} \simeq 2.5 \frac{c_\text{s0}^3}{G \Omega_\text{K0}} \left( \frac{Q_0}{\widetilde{h}_\text{p}} \right)^{-5/13}.
\end{equation}
We then find our fiducial planet mass $M_\text{p} \sim 0.15 (1 + Z)^{-3/2} M_\text{F}$. Thus we do not expect disc-feedback to play a dominant role in our simulations.

We will find that early on much of the dust-gas dynamics occur near the planet's co-orbital region. In the pure gas limit, the half-width of the horseshoe region $x_s$ is given by  
\begin{equation}
	\frac{x_s}{r_\text{p}} = 1.1 \left( \frac{0.4}{b} \right)^{\frac{1}{4}} \sqrt{\frac{q}{h_\text{p}}}\label{paar_xs}
\end{equation}
\citep{Paardekooper2010}. In a dusty disc we expect to replace $b\to r_\text{s}/\widetilde{H}_\text{p}$ and $h_\text{p}\to \widetilde{h}_\text{p}$, for which we choose the same initial values across all simulations. 
For our fiducial planet mass, $x_s\simeq 0.01r_\text{p}\simeq 0.2\widetilde{H}_\text{p}$ and the corresponding libration time-scale, $\tau_\text{lib} = 8 \pi r_\text{p} / (3 \Omega_\text{p} x_s)$, is about $120$ orbits.

\subsection{Numerical Setup}

We simulate the above disc-planet system using the \textsc{fargo3d} code with the FARGO algorithm \citep{Masset2000, BenitezLlambay2016, BenitezLlambay2019}. 
\textsc{fargo3d} evolves the hydrodynamic equations using a finite-difference method and the planet's equation of motion with a fifth-order Runge--Kutta integration. We run the code on Graphics Processing Units (GPUs), which provides significant speed-ups to allow for high-resolution, long-term simulations. The simulation domain is $(r, \phi) \in [0.4, 2.0] \times [0, 2 \pi]$. We adopt a standard resolution of $(N_r, N_\phi) = (2880, 5344)$, where cells are uniformly spaced in the domain. The effective scale height is then resolved by $(90, 40)$ cells in the radial and azimuthal direction in the vicinity of planet,  respectively. We apply periodic boundaries in $\phi$ and set reflective radial boundaries with values of surface density and azimuthal velocity extrapolated from active cells in order to preserve rotational equilibrium. We also apply damping zones in $r\in [0.40, 0.44]$ and $r\in [1.82, 2.00]$ to reduce reflections  \citep{valborro06}. Code units are such that $G=M_\star=r_0=1$. The orbital period at $r_0$ is thus $2\pi$. 

In Appendix \ref{pluto_compare} we check that our setup with \textsc{fargo3d} produces similar results to our previous work \citep{Chen2018} that employed a fundamentally different dust model and numerical code.  

\subsection{Torque Analyses}

We measure disc-on-planet torques to interpret the migration behaviour observed in the simulations. We compare the torque obtained from simulations to a modified torque formula in \citet{Chen2018}, which is based on a semi-analytic formula derived for pure-gas discs by \citet{Paardekooper2010}, but accounts for the effects of dust on the disc structure and sound-speed. For a dusty disc with constant metallicity, the disc-on-planet torque is given by 
\begin{align}
\label{eq:torque_formula}
    \frac{\Gamma}{\Gamma_\text{ref}} 
        =& ~\Gamma_\text{L} + \Gamma_\text{c} + \Gamma_\text{hs} \nonumber\\
        =& -(2.5 - 0.5 \beta - 0.1 \sigma) \left( \frac{0.4}{\widetilde{b}} \right)^{0.71} \nonumber\\
         & -1.4 \beta \left( \frac{0.4}{\widetilde{b}} \right)^{1.26} 
           + 1.1 \left( \frac{3}{2} - \sigma \right) \left( \frac{0.4}{\widetilde{b}}\right),
\end{align}
where $\Gamma_\text{L}$ is the linear Lindblad torque, $\Gamma_\text{c}$ is the linear entropy-related corotation torque, and $\Gamma_\text{hs}$ is the non-linear, potential vorticity (PV) related horseshoe drag. Here, the torque scaling is 
\begin{equation}
\label{eq:torque_ref}
\Gamma_\text{ref} = \left( \Sigma_\text{g0} + \Sigma_\text{d0} \right) 
			        r_0^4 \Omega_\text{K0}^2 
        \left( \frac{q}{\widetilde{h}_0} \right)^2. 
\end{equation}
Note that the corotation torque $\Gamma_\text{c}$ vanishes in our fiducial, isothermal discs with $\beta = 0$. In \S\ref{constPV} we consider a run with constant PV with $\sigma=1.5$, where the horseshoe drag $\Gamma_\text{hs}$ vanishes as well.

\section{Results}\label{results}

We consider discs with metallicity $Z\in[0.01, 0.1, 0.3, 0.5, 1]$ and Stokes number $\st\in[10^{-3},10^{-2},0.03,0.06,0.1]$. The $2 M_\oplus$ planet corresponds to the range from $0.15 M_\text{F}$ ($Z = 0.01$) to $0.05 M_\text{F}$ ($Z = 1$). The simulation time is 1000 orbits.

\begin{figure*}
	\includegraphics[width=2\columnwidth]{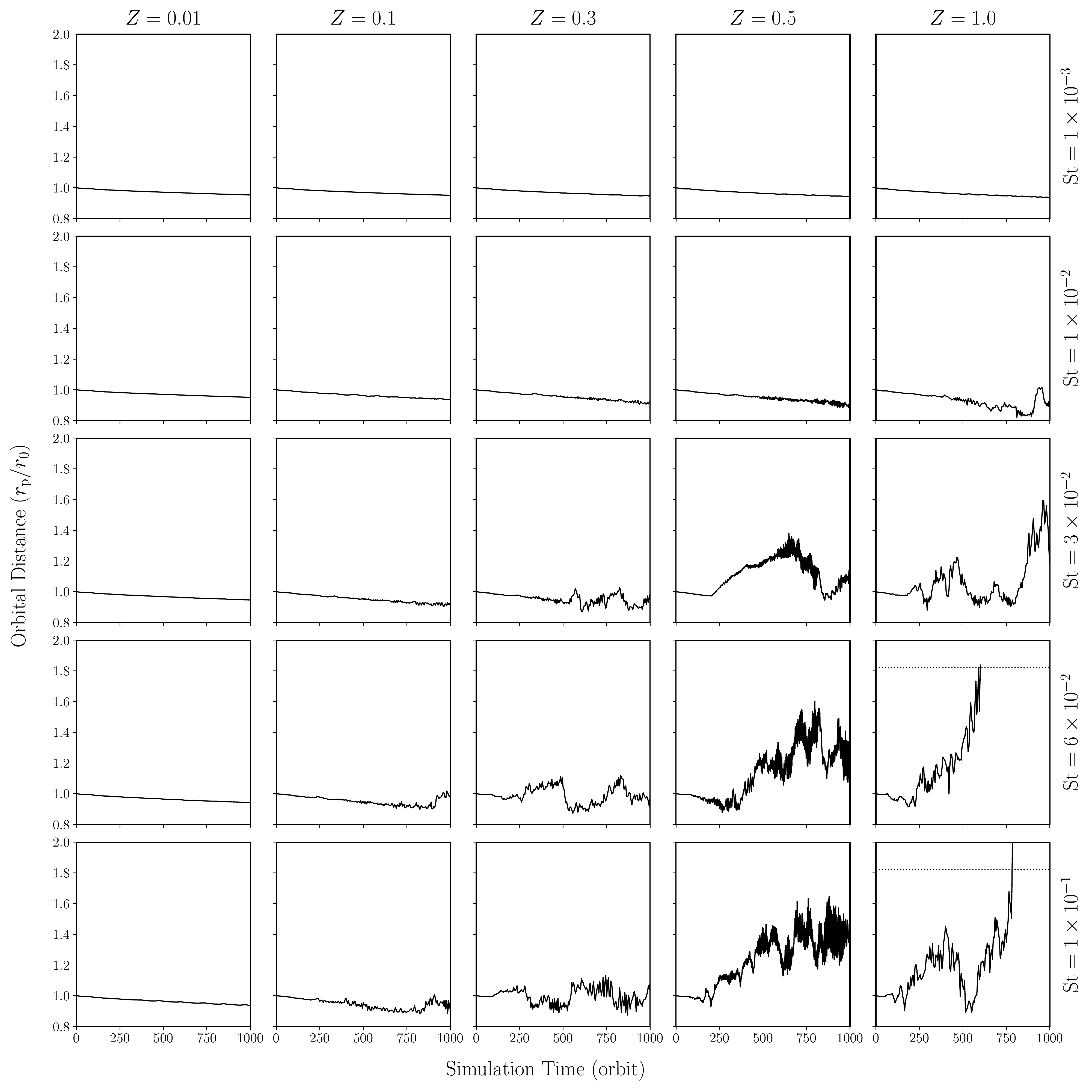}
    \caption{Orbital evolution of a $2 M_\odot$ planet in the fiducial disc 
    with different metallicities $Z$ and Stokes numbers $\st$, which increases from left to right and top to bottom, respectively. 
    The dotted lines in the bottom-right panels mark the boundary of outer damping zone. Curves are terminated once the planet reaches the damping zones, after which the simulation cannot be trusted.}
    \label{fig:Various_St_Z_orbdist}
\end{figure*}

Fig.~\ref{fig:Various_St_Z_orbdist} shows the orbital evolution of a $2 M_\oplus$ planet in the above disc models. 
The metallicity increases from left to right, and the Stokes number increases from top to bottom. In the parameter space we explored, the migration behaviour can be grouped in two categories: steady cases and stochastic cases. For discs with low metallicity ($Z \lesssim 0.1$) or small Stokes number ($\st \lesssim 10^{-2}$), the planet migrates inward steadily. When both the metallicity and Stokes number increase beyond moderate values, $Z \ge 0.3$ and $\st \ge 3 \times 10^{-2}$, the migration behaviour becomes chaotic in a few hundreds of orbits. Migration can reverse direction suddenly, making its behaviour unpredictable.
We note that in certain cases, the migration behaviour can transit from the steady case to the stochastic case at the late stage of simulations, as in the discs with $Z = 1$ and $\st = 10^{-2}$, and $Z = 0.1$ and $\st \ge 6 \times 10^{-2}$.

In this section, we first take an overview of the disc morphology with an embedded, migrating planet. We then proceed to discuss its relation to the migration behaviour in the steady and stochastic cases.

\begin{figure*}
	\includegraphics[width=2\columnwidth]{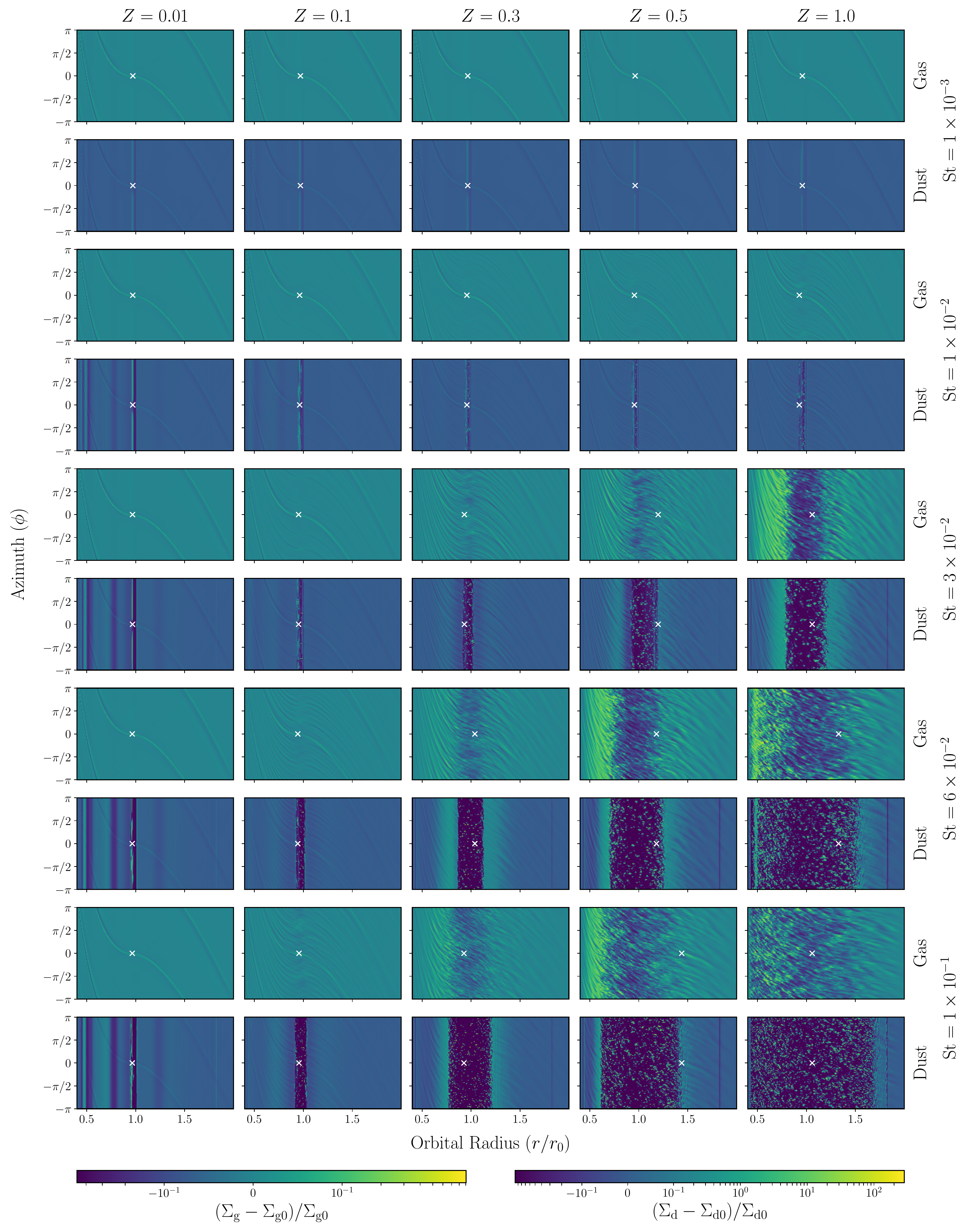}
    \caption{Relative surface density perturbation in the gas and 
    dust disc at 500 orbits. 
    For each model, the gas and dust are shown on the top and bottom in a double panel, respectively. 
    The planet position is marked by the white cross. }
    \label{fig:Various_St_Z_dens}
\end{figure*}

\begin{figure*}
	\includegraphics[width=2\columnwidth]{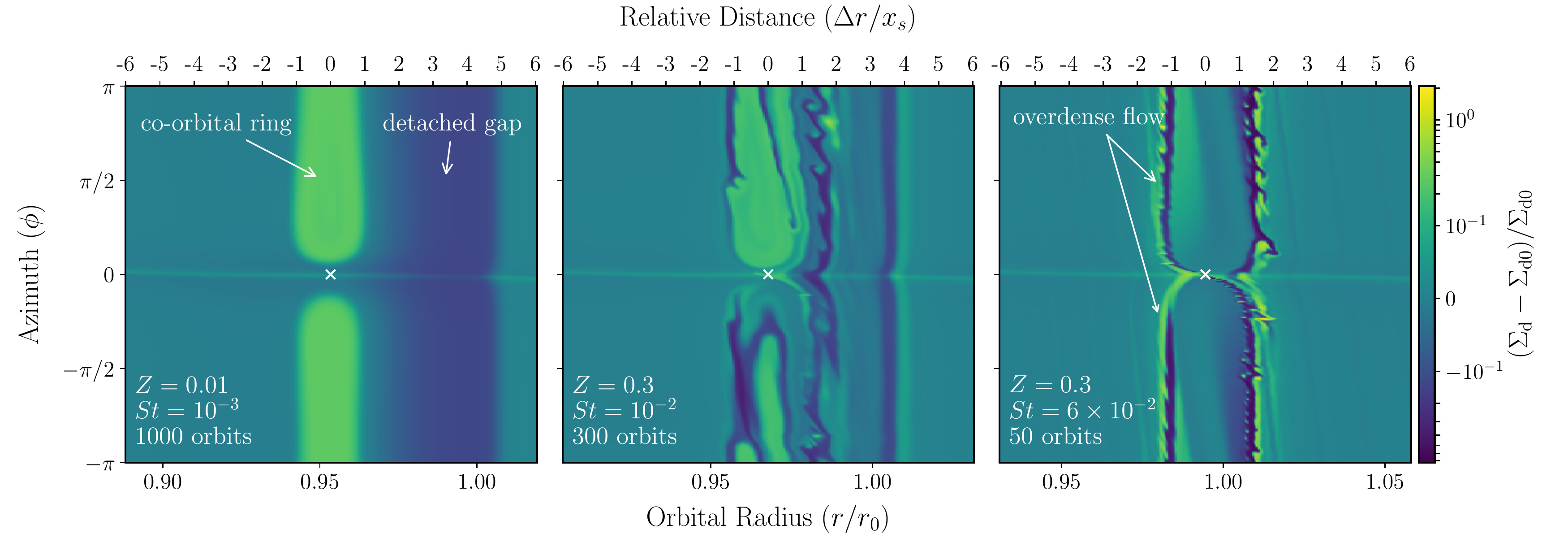}
    \caption{Relative dust surface density perturbation for disc models with 
    $Z = 0.01$ and $\st = 10^{-3}$ at 1000 orbits (left),
    $Z = 0.3$  and $\st = 10^{-2}$ at 300 orbits (middle), and
    $Z = 0.3$  and $\st = 6 \times 10^{-2}$ at 50 orbits (right).
    The upper horizontal axis shows the distance relative to the migrating planet, $\Delta r \equiv r - r_\text{p}$,
    in units of half-width of horseshoe region $x_s$.}
    \label{fig:Various_St_Z_dens_char}
\end{figure*}

\begin{figure}
	\includegraphics[width=\columnwidth]{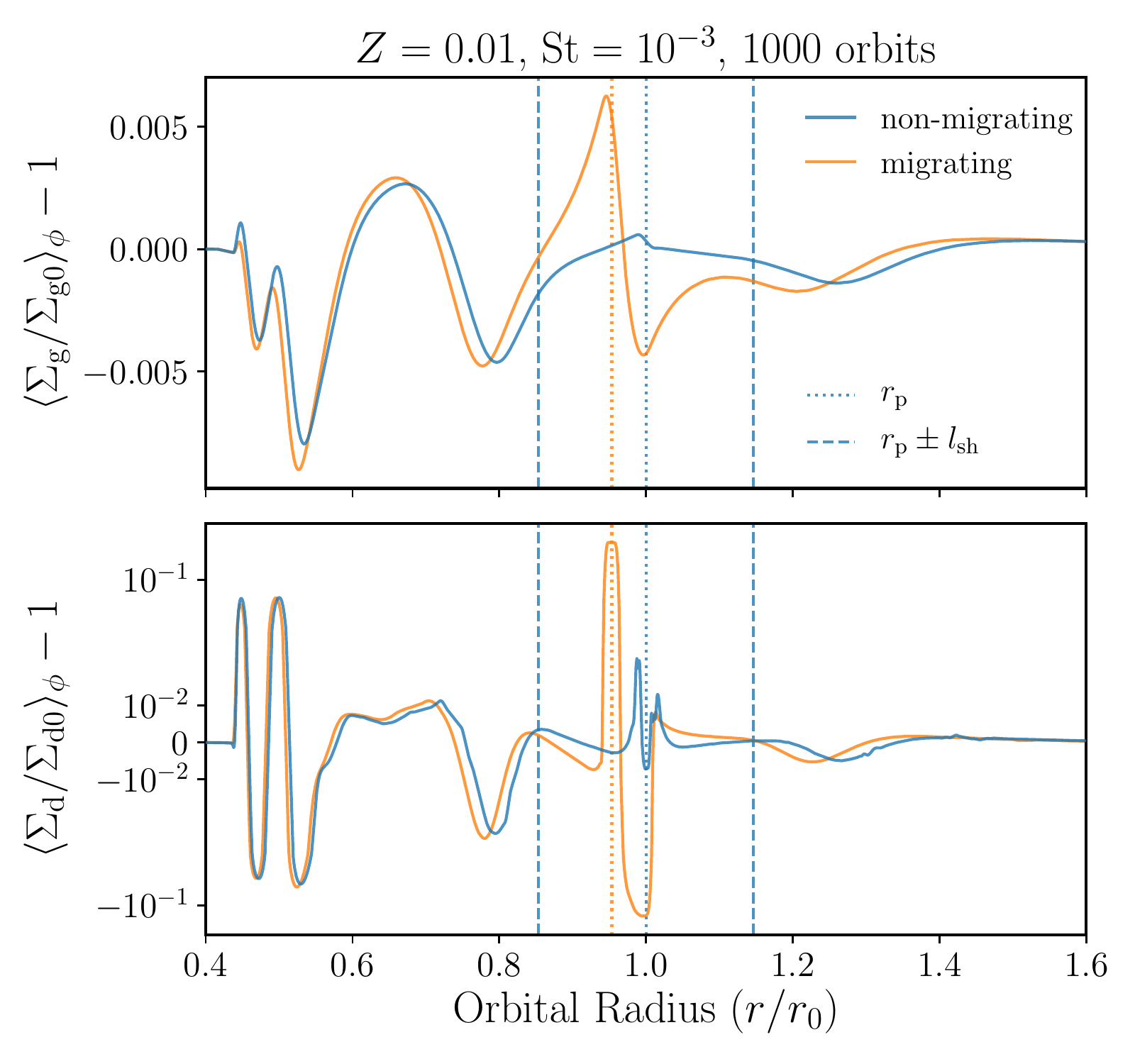}
    \caption{Azimuthally averaged, relative surface density perturbation in the gas disc (top) and dust disc (bottom)
     at $1000$ orbits for the model with $Z = 0.01$ and $\st = 10^{-3}$. The blue and orange curves corresponds to non-migrating and migrating planets, respectively. 
     The vertical dotted lines mark the planet position 
     and the vertical dashed lines mark the estimated shock location 
     of planet-induced spirals for the non-migrating planet. 
    \label{fig:St1e-3_Z0.01_densave_nonmig_mig}}
\end{figure}

\begin{figure}
	\includegraphics[width=\columnwidth]{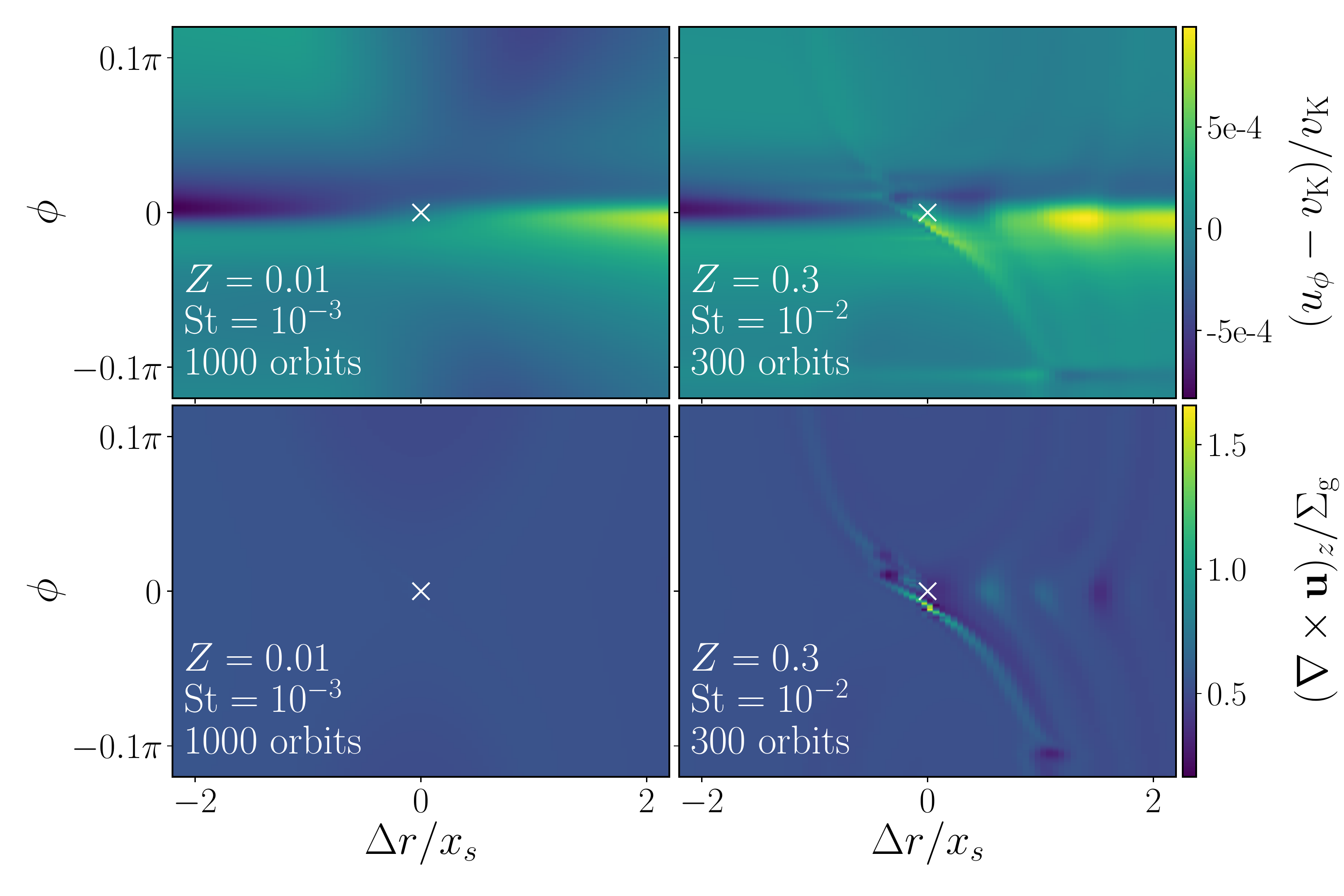}
    \caption{Top: Normalized gas azimuthal velocity in the vicinity of the migrating planet 
    for the model with 
    $Z = 0.01$ and $\st = 10^{-3}$ at 1000 orbits (left), and
    $Z = 0.3$  and $\st = 10^{-2}$ at 300 orbits (right). Here $v_\text{K} = r\Omega_\text{K}$ is the Keplerian velocity.
    Bottom: Corresponding potential vorticity. }
    \label{fig:Two_StZ_vtg_PV}
\end{figure}

\subsection{Disc Morphology}

Fig.~\ref{fig:Various_St_Z_dens} show snapshots of the surface densities at 500 orbits.
The gas and dust discs are shown on the top and bottom in a double panel, respectively. 
Depending on the metallicity and Stokes number, different substructure features are seen in the dust disc: multiple rings and gaps in discs with low metallicity or small Stokes number, and a wide gap with numerous small-scale dust vortices in discs with high metallicity and large Stokes number.
In Fig.~\ref{fig:Various_St_Z_dens_char} we re-plot the dust surface density in the vicinity of planet for three models at different simulation times, which presents the characteristic stage of dust disc evolution in our simulation set.

We first focus on the model with $Z = 0.01$ and $\st = 10^{-3}$.
Previous hydrodynamic simulations have shown that in a low-viscosity discs, a low-mass planet can induce multiple rings and gaps in both the gas and dust discs \citep{Zhu2014, Dong2017, Dong2018}.
The shocking and dissipation of density wakes leads to gap formation in the gas disc \citep{Goodman2001, Rafikov2002a, Rafikov2002b}. Dust particles then drift towards the pressure maxima at gap edges due to aerodynamic drag, resulting in the formation of dust gaps/rings. We demonstrate this with a non-migrating run and plot the surface densities in Fig.~\ref{fig:St1e-3_Z0.01_densave_nonmig_mig} (blue curves). The double gap at $r = 0.8r_0$ and $1.2r_0$ are opened by the dissipation of the primary density wakes located away from the planet at distances of \citep[see e.g.][]{Dong2017}
\begin{equation}
	l_\text{sh} \sim 0.8 \left( \frac{\gamma + 1}{12 / 5} 
	                            \frac{M_\text{p}}{M_\text{th}} \right)^{-2 / 5} H \label{lshock},
\end{equation}
where $M_\text{th} \equiv c_\text{s}^3 / (G \Omega_\text{K})$ is the thermal mass\footnote{
Depending on the metallicity, a $2M_\oplus$ planet in our fiducial model corresponds to the range from $0.05 M_\text{th}$ ($Z = 0.01$) to $0.02 M_\text{th}$ ($Z = 1$).
} and $\gamma$ is the adiabatic index ($=1$ in our case). 
The two gaps opened interior to $r = 0.6r_0$ result from the dissipation of additional density wakes excited by the planet \citep{Bae2017}. These 4 dust gaps, opened by the action of Lindblad torques, are deeper and more visible with larger Stokes numbers as a result of larger drift velocities (see Fig.~\ref{fig:Various_St_Z_dens}). In addition, there is a narrower and shallower dust gap in the vicinity of the planet. This differs from non-migrating planets embedded in high-metallicity discs, in which \citet{Pierens2019} found that a double dust gap is formed at a distance of $1 - 2 x_s$ on either side of the planet.

When the planet is allowed to migrate, a gas surface density (and hence pressure) bump develops in front of the planet, as shown by the orange curve in the top panel of Fig.~\ref{fig:St1e-3_Z0.01_densave_nonmig_mig}.
This results from the pile-up of gas pushed inwardly by the migrating planet \citep{Fung2017}.
Consequently, dust particles accumulate in the vicinity of planet, leading to the formation of a co-orbital dust ring. 
In addition, the back-reaction from drifting particles just outside the planet causes the gas to flow outwards, resulting in a density minimum located near the initial position of the embedded planet.
This leads to a dust gap that develops just outside the planet's orbital radius, but it does not follow the migrating planet inwards. In this sense the dust gap is `detached' and widens with time. The co-orbital dust ring and detached gap, as illustrated in the left panel of Fig.~\ref{fig:Various_St_Z_dens_char}, are the common features seen in our simulations for discs with $Z \le 0.1$ or $\st \le 10^{-2}$. Similar displacements between a migrating planet and associated dust rings have been reported by \citet{Meru2019}. 

The metallicity in the dust rings
increase with time as a result of the accumulation of dust as they drift towards the local pressure maximum. As shown in the middle panel of  Fig.~\ref{fig:Various_St_Z_dens_char}, we observe that the co-orbital dust ring becomes unstable. We find that in discs with $\st \ge 10^{-2}$ dust vortices are generated from the unstable co-orbital dust ring \citep[see also][]{Yang2020}. 

The above phenomenon is consistent with the recent study from \citet{Huang2020}. They find that when the metallicity in a dust ring is of order unity with a sharp density contrast at the ring edges, dust feedback can alter the gas azimuthal velocity that enhances the local gradient of potential vorticity (PV), which then leads to meso-scale instabilities and the formation of small-scale, dust vortices. The instability may be a dusty analog of the Rossby Wave Instability (RWI) associated with PV extrema in gas discs \citep{Lovelace1999,li00}, which also leads to vortex formation \citep{li01}. 

To check for this scenario, in the bottom panels of Fig.~\ref{fig:Two_StZ_vtg_PV} we plot the potential vorticity (PV; also known as vortensity) of the gas disc, which is defined as 
\begin{equation}
    \zeta = \frac{\left( \nabla \times {\bf{u}} \right)_z }{\Sigma_\text{g}}.
\end{equation}
We find that the PV is significantly enhanced at the outer separatrix in the disc with $\st = 10^{-2}$. This results from the scattering of librating, co-orbital dust by the planet whenever it undergoes horseshoe turns, leading to the formation of overdense dust flow at the downstream separatrix \citep{Morbidelli2012, llambay18, Pierens2019}.
When the metallicity in the co-orbital ring is high, feedback from the scattered, overdense flow can significantly modify the local gas azimuthal velocity, as shown in the top panels of Fig.~\ref{fig:Two_StZ_vtg_PV}. The sharp contrast in the azimuthal velocity then contributes to the steep PV gradient that drives the RWI and eventually vortex formation.

The dust vortices that initial form in the overdense flow have radial length-scales ranging from $0.01 \widetilde{H}$ to $0.2 \widetilde{H}$; length-to-width aspect ratios 
ranging from $2$ to $6$, and masses of ranging from $10^{-7}M_\star$ to $10^{-6}M_\star$ which corresponds to $0.03 - 0.3 M_\oplus$ for a solar-mass star. The feedback from dust vortices induce pressure bumps in the gas disc, which helps to capture more dust particles, leading to the growth of dust vortices with increased metallicity. For vortices in discs with larger Stokes number, the growth rate is higher due to the larger particle drift velocities. Dust vortices with lower metallicity tend to drift faster in the azimuthal direction. 
As a result, dust vortices with different metallicities can catch up with one another and  merge to form larger vortices. 

In discs with high metallicity and large Stokes number, $Z \ge 0.3$ and $\st \ge 3 \times 10^{-2}$, we find dust vortices are quickly formed at the overdense flow 
before the co-orbital dust ring is fully developed, as illustrated in the right panel of Fig.~\ref{fig:Various_St_Z_dens_char}. These vortices grow and clear out the surrounding material to form a deep dust gap filled with dust vortices. The dust rings at the gap edges collect more particles and eventually the large gradients trigger RWI-type instabilities, leading to the formation of new dust vortices \citep{Yang2020} that clear more material. The dust gap thus widens and deepens with time \citep[see also][]{Pierens2019}.

\subsection{Steady Migration}

The previous section showed that different substructures develop in discs with different metallicities and Stokes numbers. We now investigate the effects of these substructures on disc-planet torques and the orbital migration of the embedded planet. We first describe cases with steady migration, which occurs in discs with low metallicity and small Stokes number, $Z \lesssim 0.1$ and $\st \lesssim 10^{-2}$. In this limit a co-orbital dust ring and a detached dust gap are formed in the vicinity of planet, with no or a few dust vortices generated. We find that orbital migration can oscillate on short time-scales, but overall the planet still migrates inwards.

\begin{figure}
	\includegraphics[width=\columnwidth]{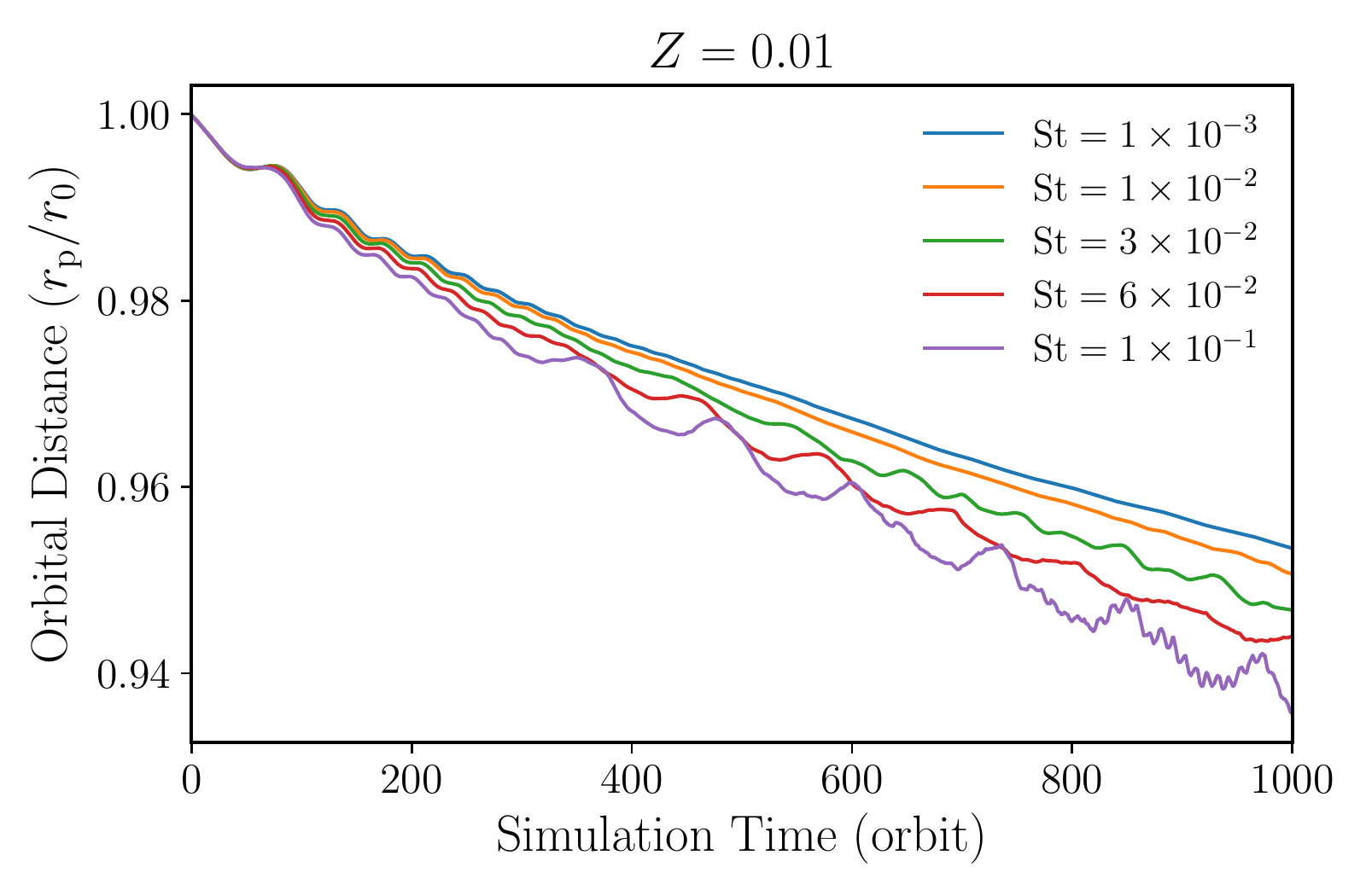}
    \caption{Orbital evolution of a $2 M_\oplus$ planet in a dusty disc with $Z = 0.01$ 
    and different Stokes numbers.}
    \label{fig:Steady_Z0.01_St_rp}
\end{figure}

\begin{figure}
	\includegraphics[width=\columnwidth]{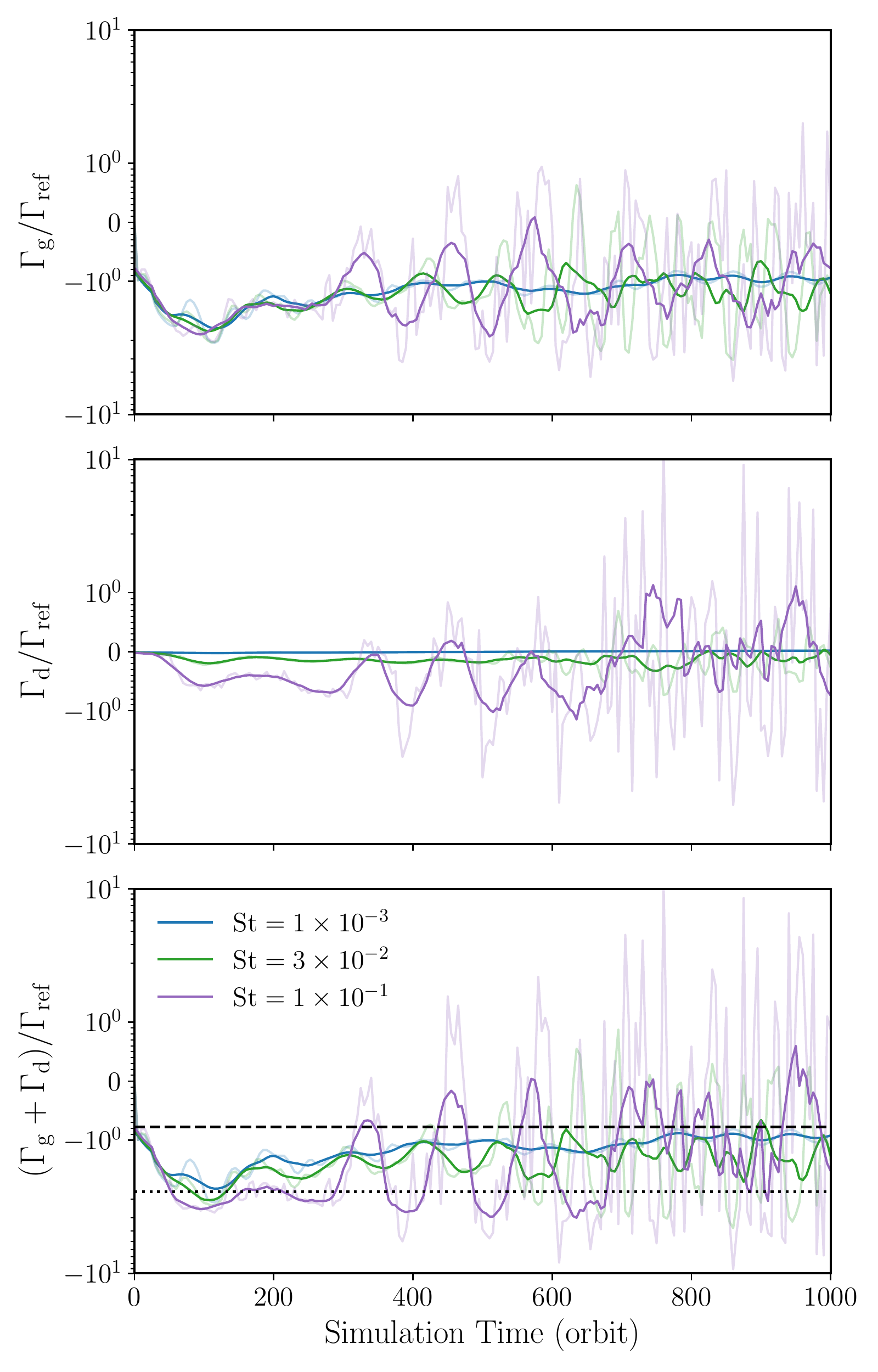}
    \caption{Evolution of torque exerted on the planet by the gas disc (top), the dust disc (middle),
    and their sum (bottom). Light coloured lines show the instantaneous torque, 
    and dark coloured lines show 50-orbit running averages.
    In the bottom panel, the dotted and dashed lines are the semi-analytical values 
    of Lindblad torque and total disc-on-planet torque  
    obtained from equation~(\ref{eq:torque_formula}), respectively.}
    \label{fig:Steady_Z0.01_St_torque}
\end{figure}

In Fig.~\ref{fig:Steady_Z0.01_St_rp} we compare the planet's orbital evolution in discs with $Z = 0.01$ and varying Stokes numbers, which all show that the average migration rate slows down over time. (Similar results are obtained for fixed Stokes numbers and increasing metallicities.) This is due to dynamical corotation torques acting on migrating planets \citep{Paardekooper2014}. In our fiducial disc with negative radial PV gradients, the gas dynamical torque is positive for inwards planet migration. Fig.~\ref{fig:Steady_Z0.01_St_torque} shows the evolution of disc-on-planet torque from the gas disc, the dust disc, and their sum. We see that both the gas and dust torques become saturated at 100 orbits, but the gas torque becomes more positive due to the dynamical torque.  

\begin{figure}
	\includegraphics[width=\columnwidth]{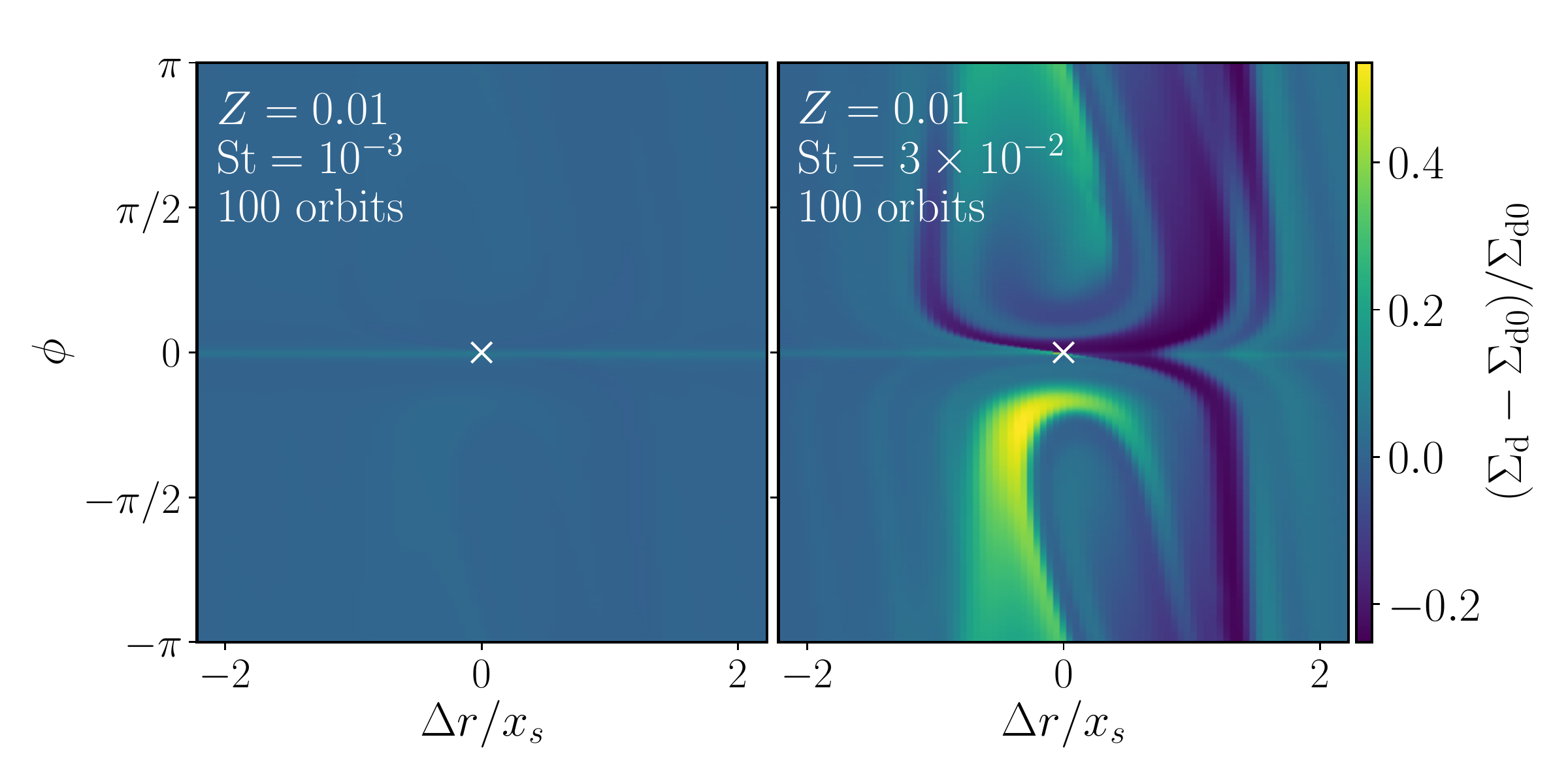}
    \caption{Relative dust surface density perturbation in the vicinity of planet at $100$ orbits
    for the model with 
    $Z = 0.01$ and $\st = 10^{-3}$ (left), and with
    $Z = 0.01$ and $\st = 3 \times 10^{-2}$ (right). Here $\Delta r \equiv r - r_\text{p}$.}
    \label{fig:Steady_dustdens}
\end{figure}

In discs with large Stokes number, the dust torque from the flow in the co-orbital region becomes important. 
Fig.~\ref{fig:Steady_dustdens} shows the normalized dust surface density and dust torque at $100$ orbits, before the co-orbital dust ring is formed, for the models with $\st = 10^{-3}$ and $\st = 3 \times 10^{-2}$. As the Stokes number increases, as in the case of $\st = 3 \times 10^{-2}$, the dust flow scattered by the planet can lead to a prominent asymmetric density distribution in the co-orbital region, which exerts a net torque on the planet \citep{llambay18}. 
 In our case this net torque is negative and results from particles scattered outwards relative to the inwardly migrating planet. The negative dust torque increases in magnitude with the Stokes number, even in discs with the same initial dust surface density (see Fig.~\ref{fig:Steady_Z0.01_St_torque}). 
The result is that the average migration rate is larger in discs with larger Stokes number, as shown in Fig.~\ref{fig:Steady_Z0.01_St_rp}. 

We find that an overdense, asymmetric co-orbital dust blob is formed in a few hundred orbits in discs with $\st \ge 3 \times 10^{-2}$. As the co-orbital dust flow librates about the planet, it introduces unsteadiness in the dust torque  \citep{Chen2018}. The feedback from the dust flow also induces periodic oscillations in the gas torque, as seen in Fig.~\ref{fig:Steady_Z0.01_St_torque}. The oscillation in the disc-on-planet torque causes the planet to waggle in the disc with a period of the libration time-scale, $\sim 120$ orbits.

In the case of $\st = 10^{-1}$, small-scale vortices are generated in the dust disc after $500$ orbits 
and the number of vortices reaches $10$ at $750$ orbits.
The dust vortices, as well as the vortex-induced spiral structures in the gas disc, produce strong variations in the disc-on-planet torque as they move past the planet. This leads to the spikes in the disc torque displayed by the light purple line in Fig.~\ref{fig:Steady_Z0.01_St_torque}, and the small-amplitude wiggle in the orbital distance shown in Fig.~\ref{fig:Steady_Z0.01_St_rp}.
We find that the effect of vortices on the migration rate is negligible if only a few small-scale vortices are present in the co-orbital region. 
However, when the number of vortices is large, they consecutively interact with the planet and significantly affect its migration behaviour, as in the stochastic case discussed in the next section.

\subsection{Stochastic Migration}

In simulations where both the metallicity and Stokes number are large ($Z \ge 0.3$ and $\st \ge 3 \times 10^{-2}$), we find more than hundreds of dust vortices are formed in the vicinity of planet before the co-orbital dust ring is developed.
When the librating dust vortices perform horseshoe turns, they exert strong torques on the planet that can cause its instantaneous migration rate to exceed $10^{-3}r_0$ per orbit, which is about an order of magnitude larger than that due to Lindblad torques. These vortex-planet scattering events dominate the migration behaviour, causing it to be chaotic and  unpredictable, even in discs with metallicity initially smaller than unity. 
However, we find the migrating planet is more or less located  within the dust gap, suggesting the planet is confined by the gap edges, as discussed below.
In this section, we first describe the run with $Z = 0.3$ and $\st = 6 \times 10^{-2}$ as an example of dust vortex-driven, stochastic  migration. 

\begin{figure}
	\includegraphics[width=\columnwidth]{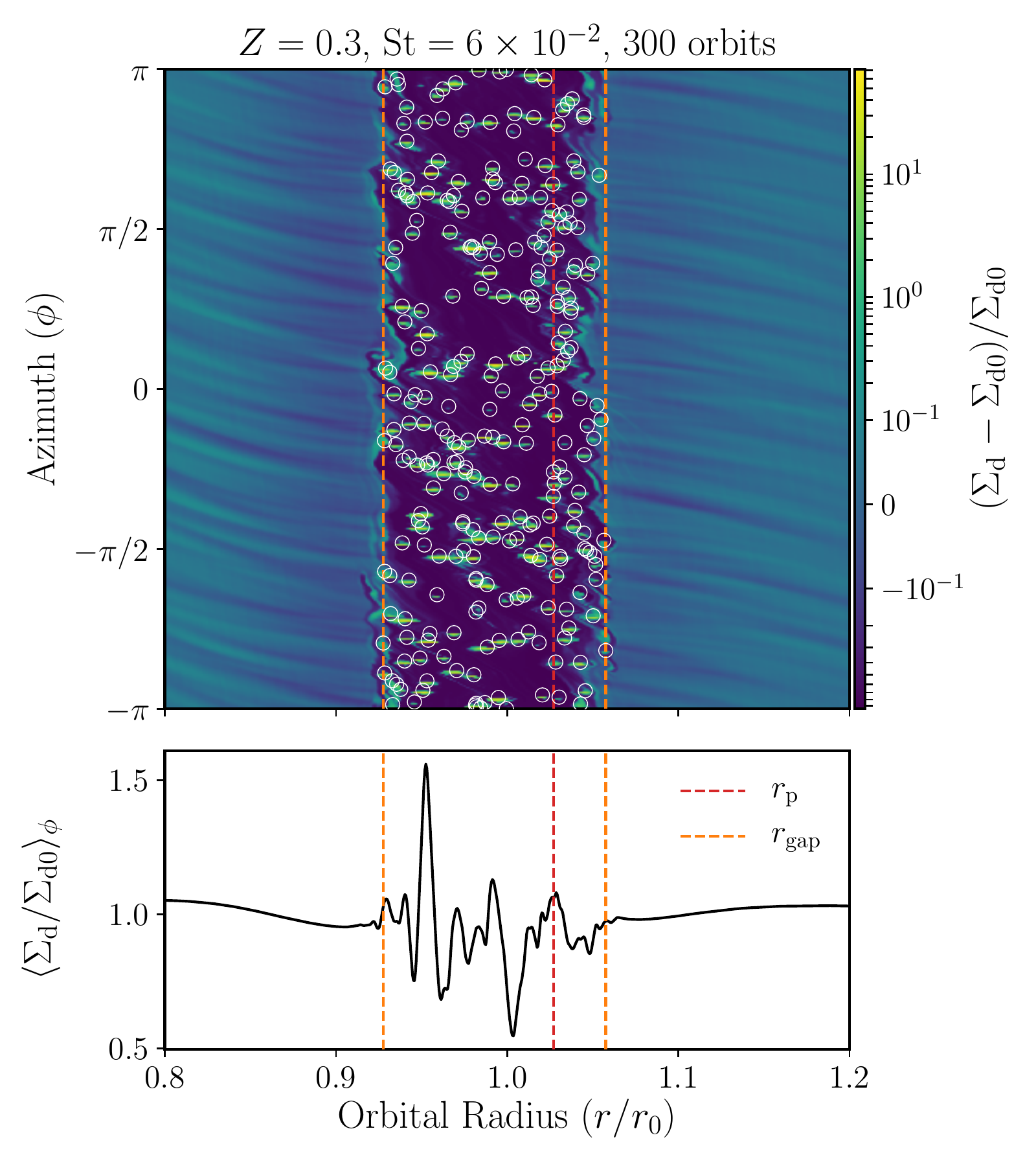}
    \caption{Top: Snapshot of normalized dust surface density at $300$ orbits 
    for the model with $Z = 0.3$ and $\st = 6 \times 10^{-2}$.
    Dust vortices are marked by white circles. 
    Bottom: Corresponding azimuthally averaged, relative dust surface density perturbation. 
    The orange vertical lines denote the location of outermost vortices and the red line marks the planet position.}
    \label{fig:Stochastic_dustgap}
\end{figure}

Fig.~\ref{fig:Stochastic_dustgap} shows the normalized dust surface density at $300$ orbits and the corresponding azimuthally-averaged profile. As shown in the bottom panel, the presence of dust vortices makes it difficult to identify a dust gap from the one-dimensional profile.  
To circumvent this, we first use the DAOFIND algorithm \citep{Stetson1987}
to identify dust vortices by searching for Gaussian-like structures in 2D maps of the dust surface density. These are marked by white circles in Fig.  \ref{fig:Stochastic_dustgap}.   
We then define the gap edge to be the location of the centroid of outermost dust vortices, as displayed by the orange lines in Fig.~\ref{fig:Stochastic_dustgap}. By inspection, this method gives an adequate estimate of the gap edges.

\begin{figure}
	\includegraphics[width=\columnwidth]{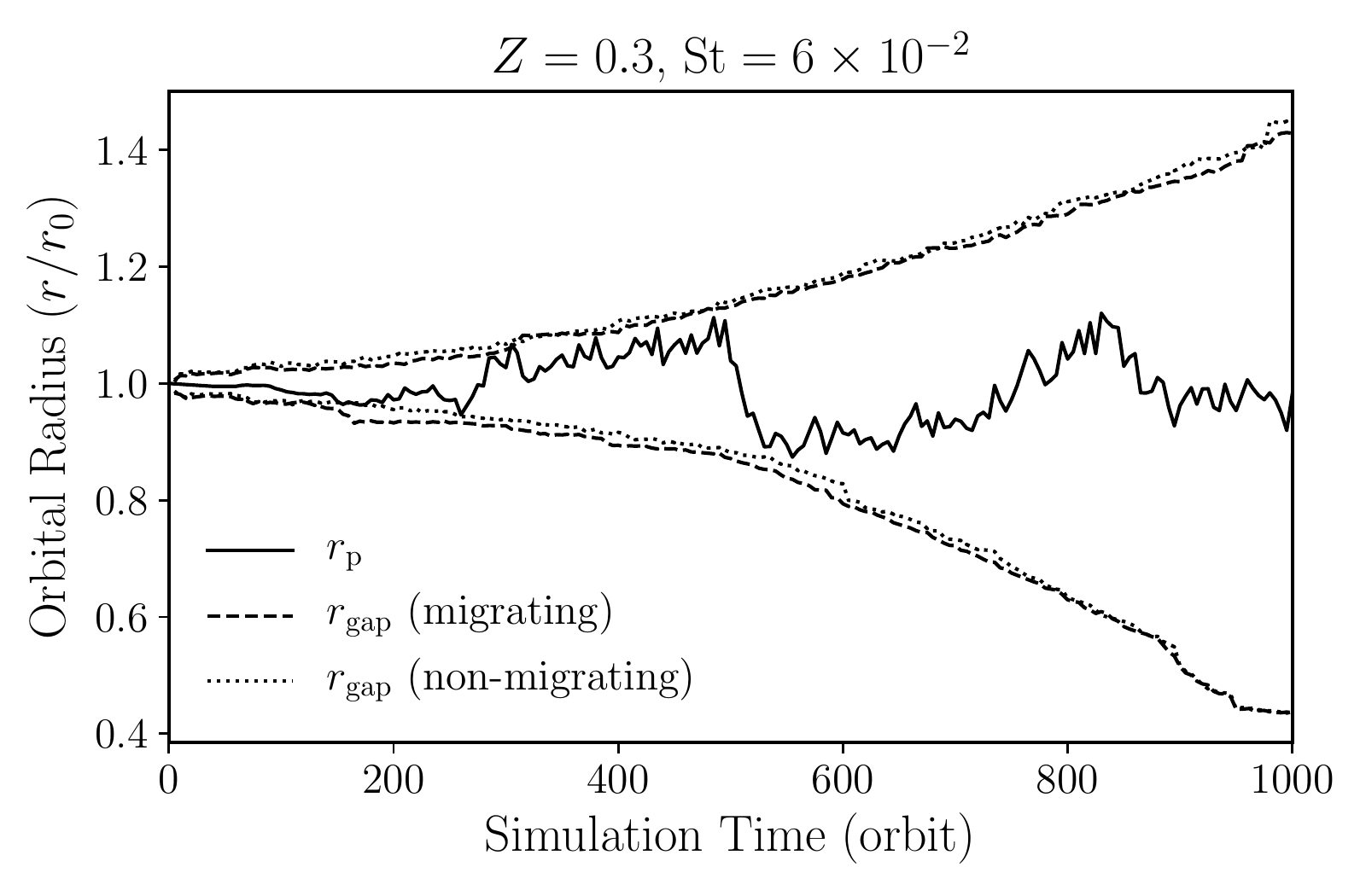}
    \caption{Evolution of orbital distance and gap edge in the disc with $Z = 0.3$ and $\st = 6 \times 10^{-2}$.
    The dashed line denotes the edge of dust gap in the migrating planet case 
    and the dotted line is the gap edge in the non-migrating planet case.    }
    \label{fig:Stochastic_rp_dustgap}
\end{figure}

\begin{figure}
	\includegraphics[width=\columnwidth]{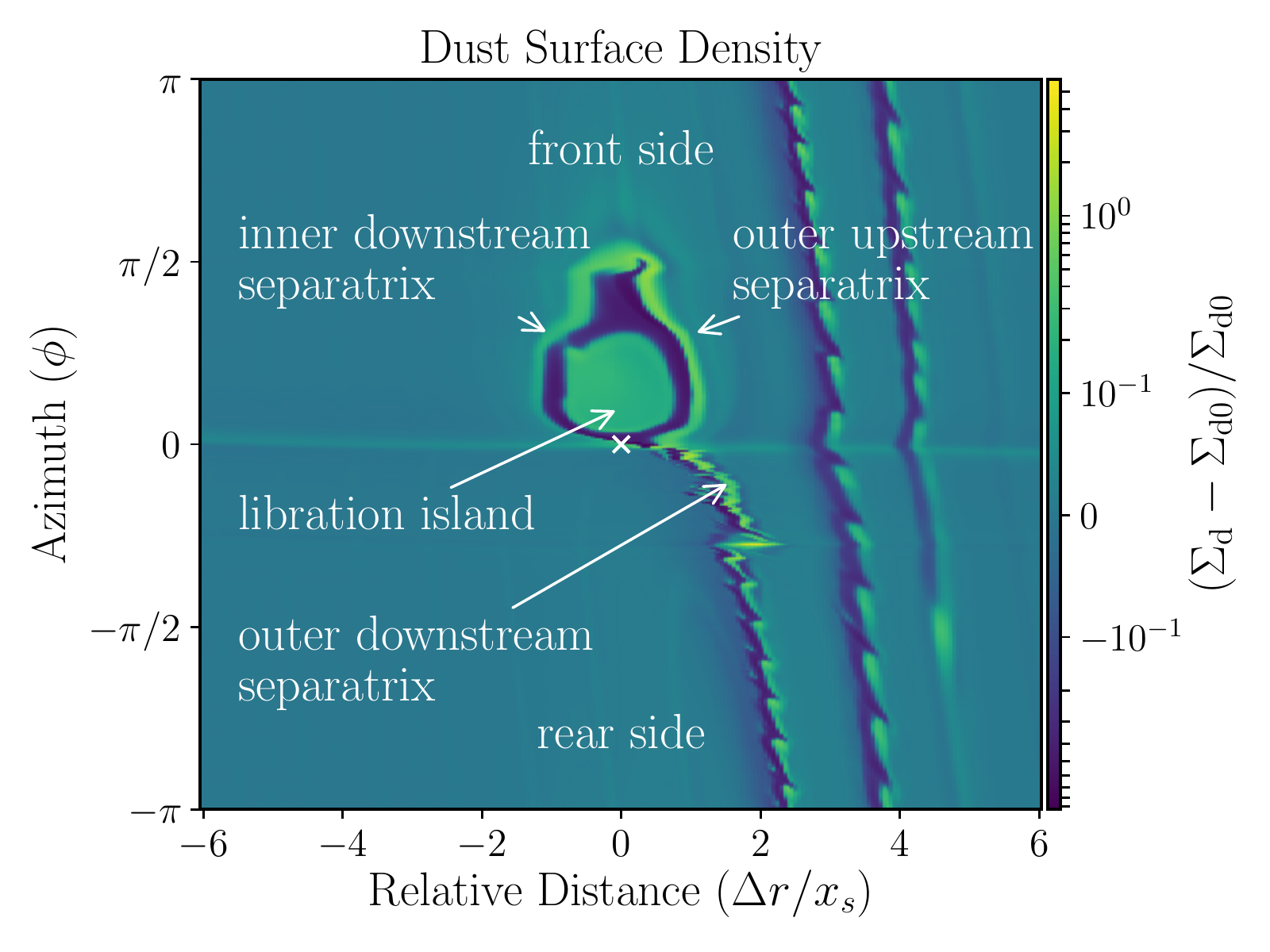}
    \caption{Schematic diagram of dusty substructures  induced by an inwardly migrating planet.  
    The planet position is marked by a white cross. Here $\Delta r \equiv r - r_\text{p}$. 
    When a planet migrates inwardly, a libration island is formed at the front side of the planet, 
    as well as two overdense scattered flows at the inner and outer downstream separatrices.
    As the overdense flow at the front side approaches to the planet from the outer upstream separatrix,
    it exerts strong positive torque on to the planet that can slow down the inwards migration rate or even reverse it.}
    \label{fig:Stochastic_schematic}
\end{figure}

Fig.~\ref{fig:Stochastic_rp_dustgap} shows the evolution in the planet's orbital radius and the location of the gap edges. We also plot the edge evolution for a corresponding simulation without migration for comparison. We can see that the migrating planet is always located within the dust gap. This results from the dynamical corotation torque from dust disc and the vortex-planet interaction as follows.  

To explain the dusty dynamical corotation torque, we performed supplementary simulations in Appendix 
\ref{append:torque_constMigRate} with prescribed, constant migration rates.  Fig.~\ref{fig:Stochastic_schematic} illustrates the dusty substructures in the vicinity of an inwardly migrating planet from such a simulation. As the planet migrates inward, a libration island is formed at the front side of planet. The overdense dust flow generated at the inner downstream separatrix librates just outside the libration island, whereas the overdense flow formed at the outer downstream separatrix leaves the co-orbtial region and circulates with the planet. 
As the librating dust flow approaches to the planet from the outer upstream separatrix, it exerts positive torque on the planet that can slow down the migration rate or even drive the planet outwards. The opposite picture develops for an outwardly migrating planet with the libration island located at the rear side of planet and the dusty dynamical torque is negative.  Consequently, when the planet is scattered out of the dust gap the exterior dust disc tends to exert an opposing torque that pushes the planet back into the dust gap.

We find that when the planet is located inside the low surface density dust gap, its migration  is dominated by discrete vortex-planet interactions, which is analogous to planetesimal-driven migration \citep[see e.g.][]{Ormel2012}. However, because of the non-uniform spatial distribution of dust vortices, planet migration is sensitive to the encounter rate and thus stochastic.
For instance, as a planet located near the outer gap edge migrates inwardly, the circulating vortices inside the planet can perform horseshoe turns at the rear side that exerts negative torque and thus increases the migration rate, which results in the rapid inwards migration at $500$ orbits shown in Fig.~\ref{fig:Stochastic_rp_dustgap}. The interior circulating vortices can also undergo small angle deflections as they encounter the planet, which exerts positive torque. The result is that the planet moves back and forth about the gap edge, as seen at $300 - 500$ orbits in Fig.~\ref{fig:Stochastic_rp_dustgap}.

Interestingly, the evolution of dust gap shows little difference between the non-migrating and migrating planets, as shown by the dotted and dashed lines in Fig.~\ref{fig:Stochastic_rp_dustgap}. 
We suspect this is because the gap widens via RWI-like instabilities associated with the sharp metallicity contrast at the edge of the dust gap. This leads to the formation of new dust vortices that pushes the gap edge further out. Planet migration has only a minor effect on this process. For example, the inwardly migrating planet slightly enhances (slows down) the widening of the inner (outer) gap edge at the first 200 orbits.

\begin{figure}
	\includegraphics[width=\columnwidth]{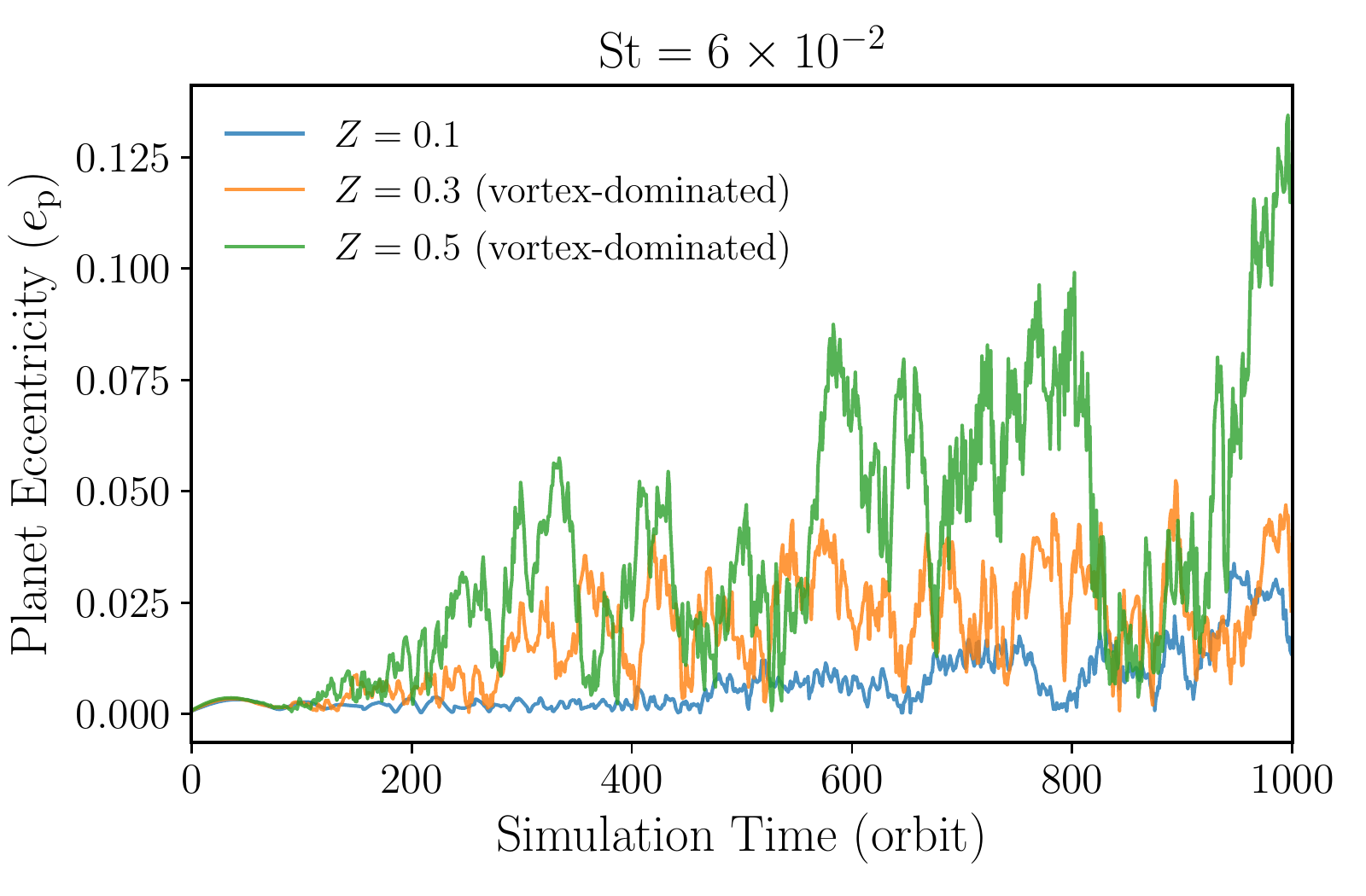}
    \caption{Evolution of planet eccentricity in discs with $\st = 6 \times 10^{-2}$ and different metallicities. Eccentricities can be excited to $O(0.1)$ in discs with vigorous vortex instabilities. }
    \label{fig:Stochastic_ep}
\end{figure}

Finally, Fig.~\ref{fig:Stochastic_ep} shows the evolution of planet eccentricity in discs with with $\st = 6 \times 10^{-2}$ and various metallicities.
We can see that compared to the model with $Z = 0.1$, vortex-planet interaction can significantly excite the planet's eccentricity.
The induced eccentricity depends on the encounter rate of dust vortices, and thus the metallicity and Stokes number. 
For discs with $Z = 0.3$ the planet eccentricity is about $0.03$, and can be higher than $0.1$ in discs with $Z \ge 0.5$.

\subsection{Effects of Viscosity and Resolution} \label{eff_vis_res}

The formation of planet-induced dust vortices via RWI-like processes (i.e. edge instabilities) can depend on resolution and viscosity  \citep{Chen2018, Pierens2019}. Here, we perform additional simulations  
with different resolutions and viscosities to study their effects on planet migration. We consider constant kinematic viscosities $\nu \in [0, 10^{-6}]$ in code units, which corresponds to alpha-viscosities of $\alpha \lesssim 3\times10^{-4}$ \citep{shakura73}. We use the same initial and boundary conditions for these low-viscosity runs as in our inviscid main runs. 

\begin{figure}
	\includegraphics[width=\columnwidth]{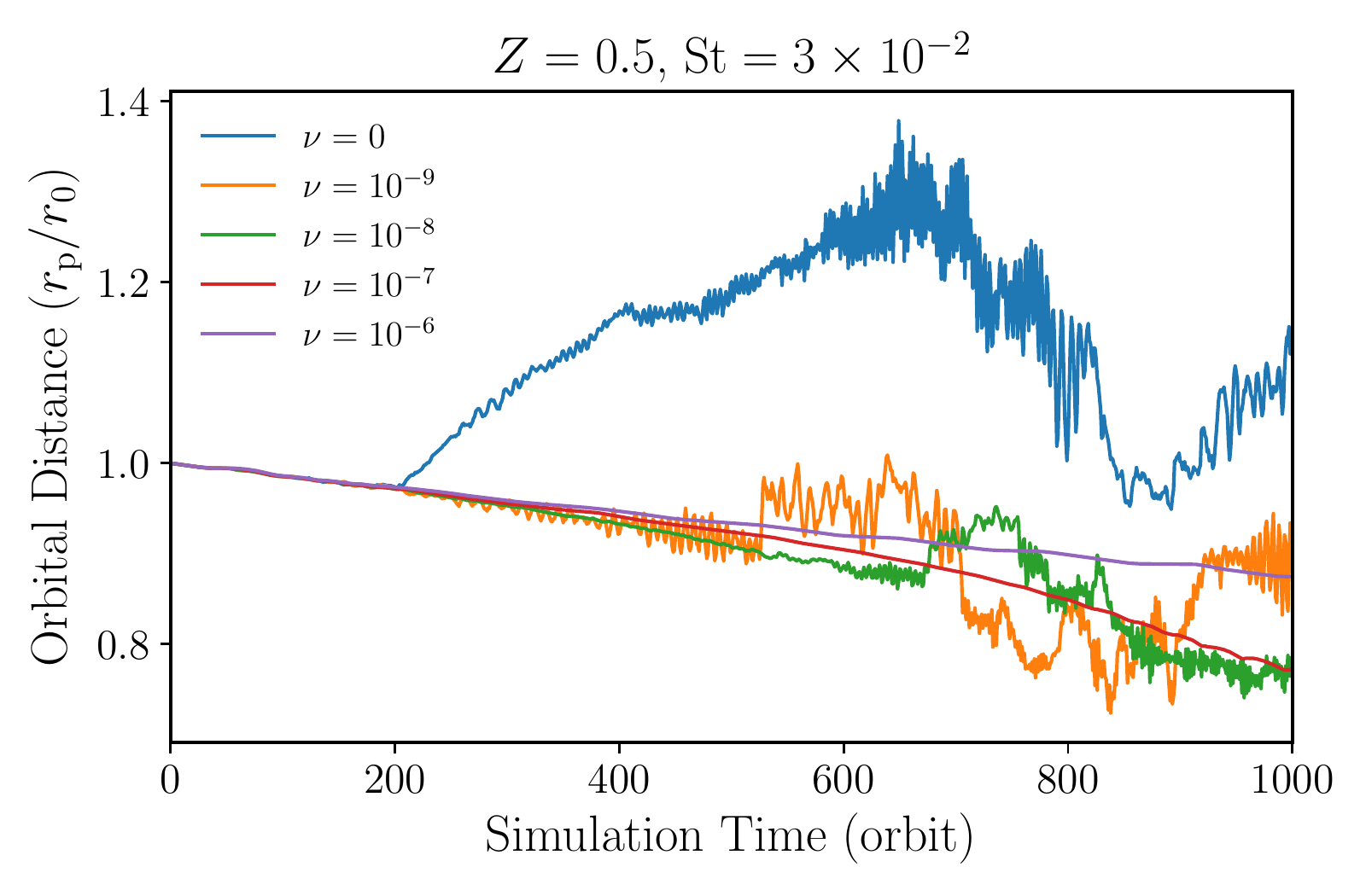}
    \caption{Orbital evolution of the migrating planet in the model with $Z = 0.5$ and $\st = 3 \times 10^{-2}$, but with different kinematic viscosity parameters $\nu$. 
    The resolution here is $(N_r, N_\phi) = (2880, 5344)$.}
    \label{fig:ConvTest_vis}
\end{figure}

Fig.~\ref{fig:ConvTest_vis} shows the orbital evolution for disc models with $Z = 0.5$ and $\st = 3 \times 10^{-2}$ and varying viscosities. 
For $\nu > 0$, the surface density of the scattered dust flow and the magnitude of the radial PV gradient near the planet separatrix are reduced, and thus the formation of dust vortices is suppressed. 
As a result, the onset of outward migration is postponed in discs with finite viscosity $0 < \nu \le 10^{-8}$. 
For higher viscosity, $\nu \ge 10^{-7}$, vortex formation is fully suppressed within the simulation time-scale. The result is that orbital migration transitions from stochastic to steady as the viscosity is increased beyond $10^{-7}$.

\begin{figure}
	\includegraphics[width=\columnwidth]{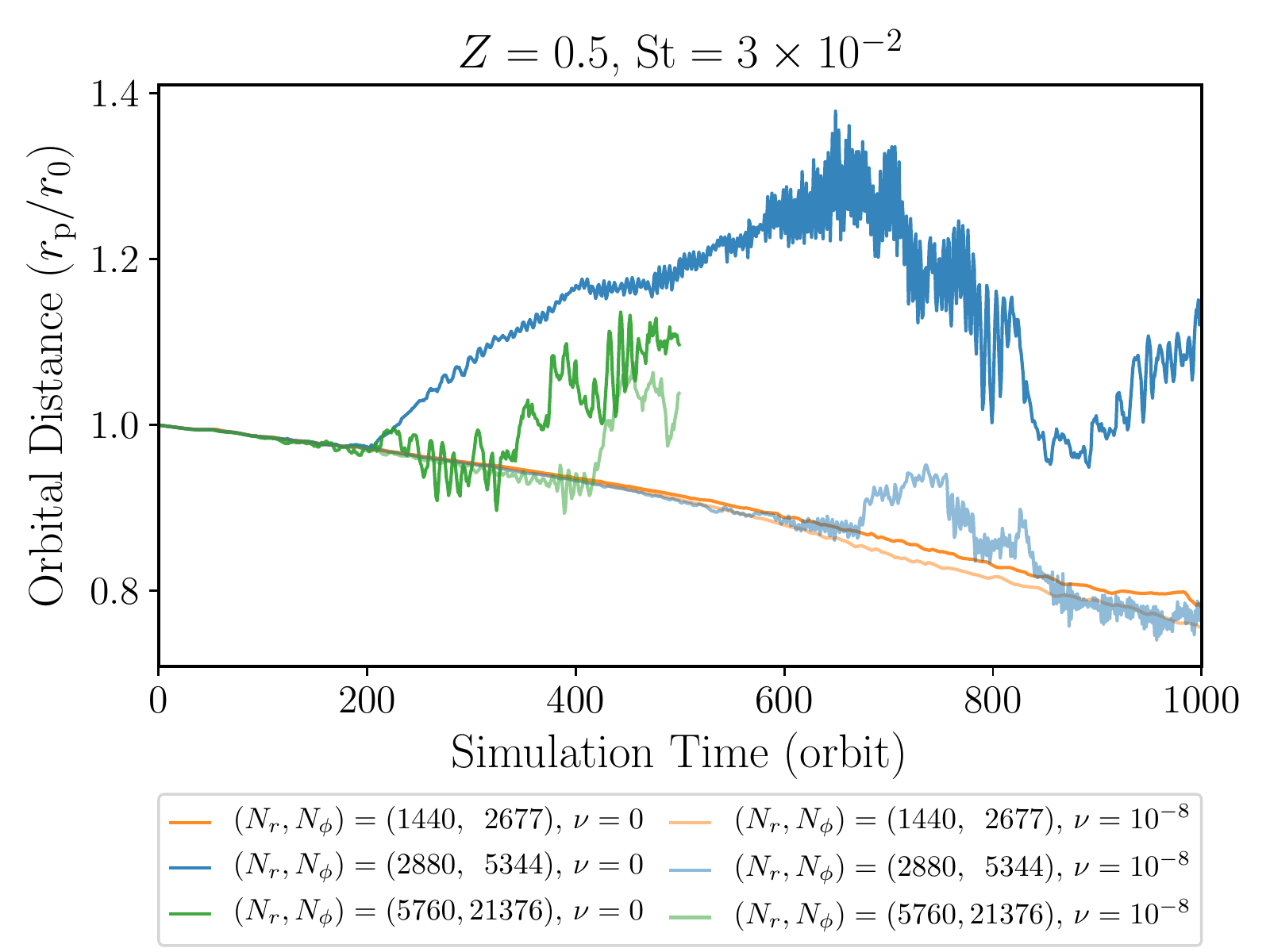}
    \caption{Same as Fig.~\ref{fig:ConvTest_vis}, but with different spatial resolutions.
    Dark coloured lines show the evolution in an inviscid disc $\nu = 0$,
    and light coloured lines are for a viscous disc with $\nu = 10^{-8}$.}
    \label{fig:ConvTest_res}
\end{figure}

Fig.~\ref{fig:ConvTest_res} shows the planet's orbital evolution, in an inviscid disc for different resolutions. Corresponding viscous runs with $\nu =10^{-8}$ are also shown for comparison. 
The resolution employed in the additional simulations are $(N_r, N_\phi) = (1440, 2677)$ and $(5760, 21376)$, where the effective scale height is resolved by $(45, 20)$ and $(180, 160)$ cells in the radial and azimuthal directions, respectively.  However, due to the increased computational cost we only simulate the  high-resolution runs up to 500 orbits.

\begin{figure}
	\includegraphics[width=\columnwidth]{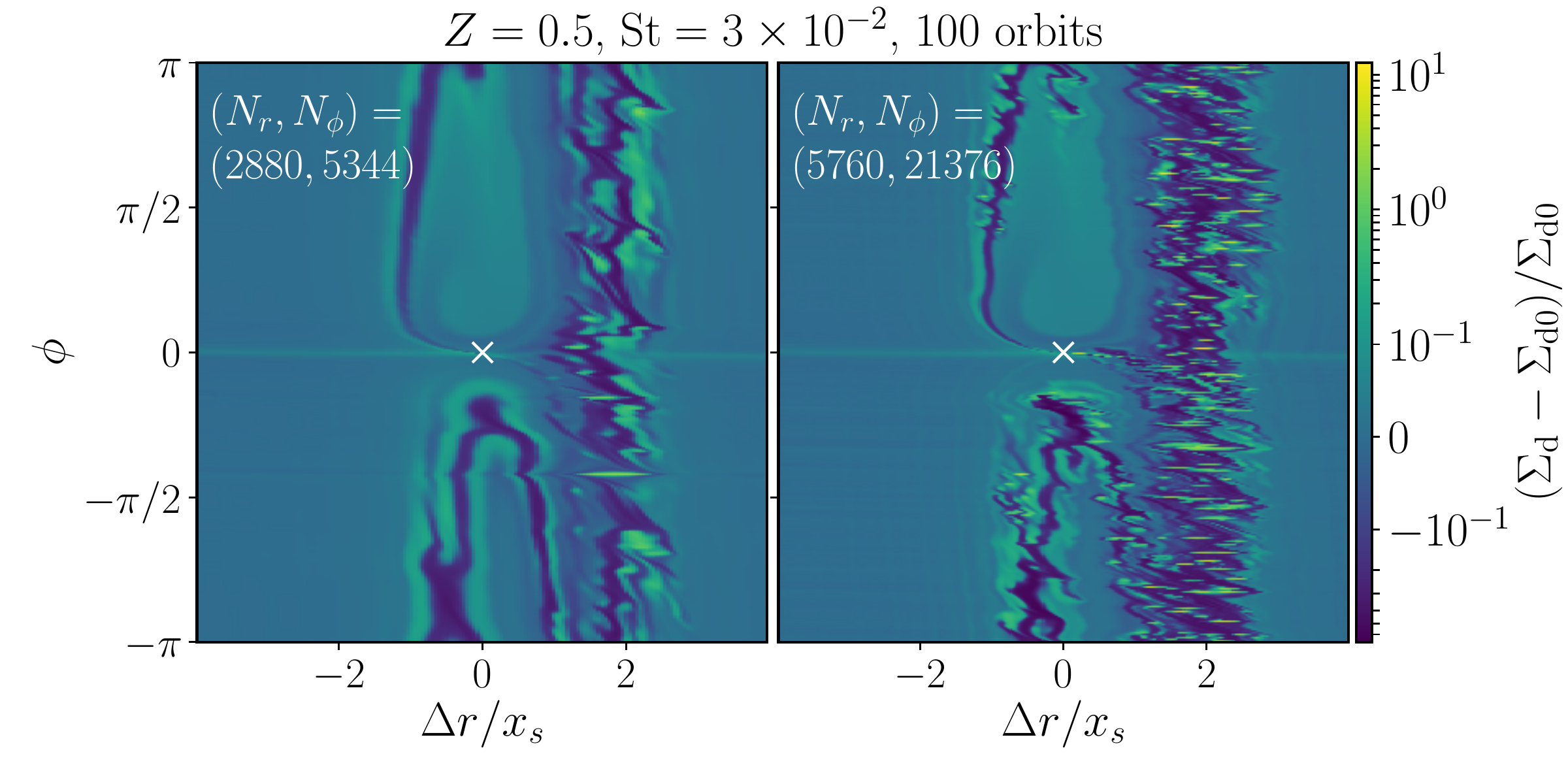}
    \caption{Relative dust surface density perturbation at $100$ orbits for the model with $Z = 0.5$ and $\st = 3 \times 10^{-2}$.
    The simulation resolution in the left panel is $(N_r, N_\phi) = (2880, 5344)$, 
    and the resolution in the right panel is $(5760, 21376)$.}
    \label{fig:ConvTest_dens}
\end{figure}

Consider first the invsicid runs. At low-resolution, vortex formation is significantly suppressed because the scattered dust flow at the downstream separatrix is not well resolved and the planet migrates inward steadily. At standard and high-resolution, the scattered dust flow is better resolved and more concentrated, which gives rise to higher metallicity and larger PV gradients that favours vortex formation. We find with increasing resolution more dust vortices with smaller radial extent are formed (see Fig.~\ref{fig:ConvTest_dens}) and planet migration becomes stochastic. 

Thus, we find that orbital migration after vortex formation ($\gtrsim 200$ orbits) does not converge with increasing resolution. However, we see in Fig. \ref{fig:ConvTest_dens} that the onset of outward migration  
converges within $200$ orbits, which is about twice the libration time-scale.
This indicates that the orbital migration in the early stages, when the disc is relatively stable, has indeed converged. 

For the viscous runs we also observe migration transitions from steady at low resolution to vortex-driven at high resolution, but outwards migration occurs earlier with increasing resolution. We also find that the effect of viscosity on vortex formation (blue and green curves) becomes less prominent with increasing resolution. 

Note that we do not include a corresponding dust diffusion in the viscous runs above \citep{youdin07}. Additional test simulations showed that dust diffusion does not affect smooth migration, nor does it prevent vortex formation and stochastic migration in low-viscosity discs. A detailed study is beyond the scope of this work, but should be considered in the future in order to have a more self-consistent physical picture.

\subsection{Effects of Initial Potential Vorticity}\label{constPV}

The initial PV profile affects disc-planet interaction and migration.  \citet{Chen2018} found that a PV blob that sustains an oscillatory corotation torque can develop in discs with non-vanishing PV gradients, but is suppressed in constant-PV discs. Furthermore, dynamical corotation torques,  which slows down inwards migration, also vanish in discs with constant PV \citep{Paardekooper2014}. We thus expect in constant-PV discs, planet torques to converge to the Lindblad values and the planet to migrate inwards faster than in non-uniform PV discs. To demonstrate this, we here consider discs with a gas surface density slope $\sigma = 1.5$ so that the initial PV ($\propto r^{3/2-\sigma}$) is nearly constant.

\begin{figure}
	\includegraphics[width=\columnwidth]{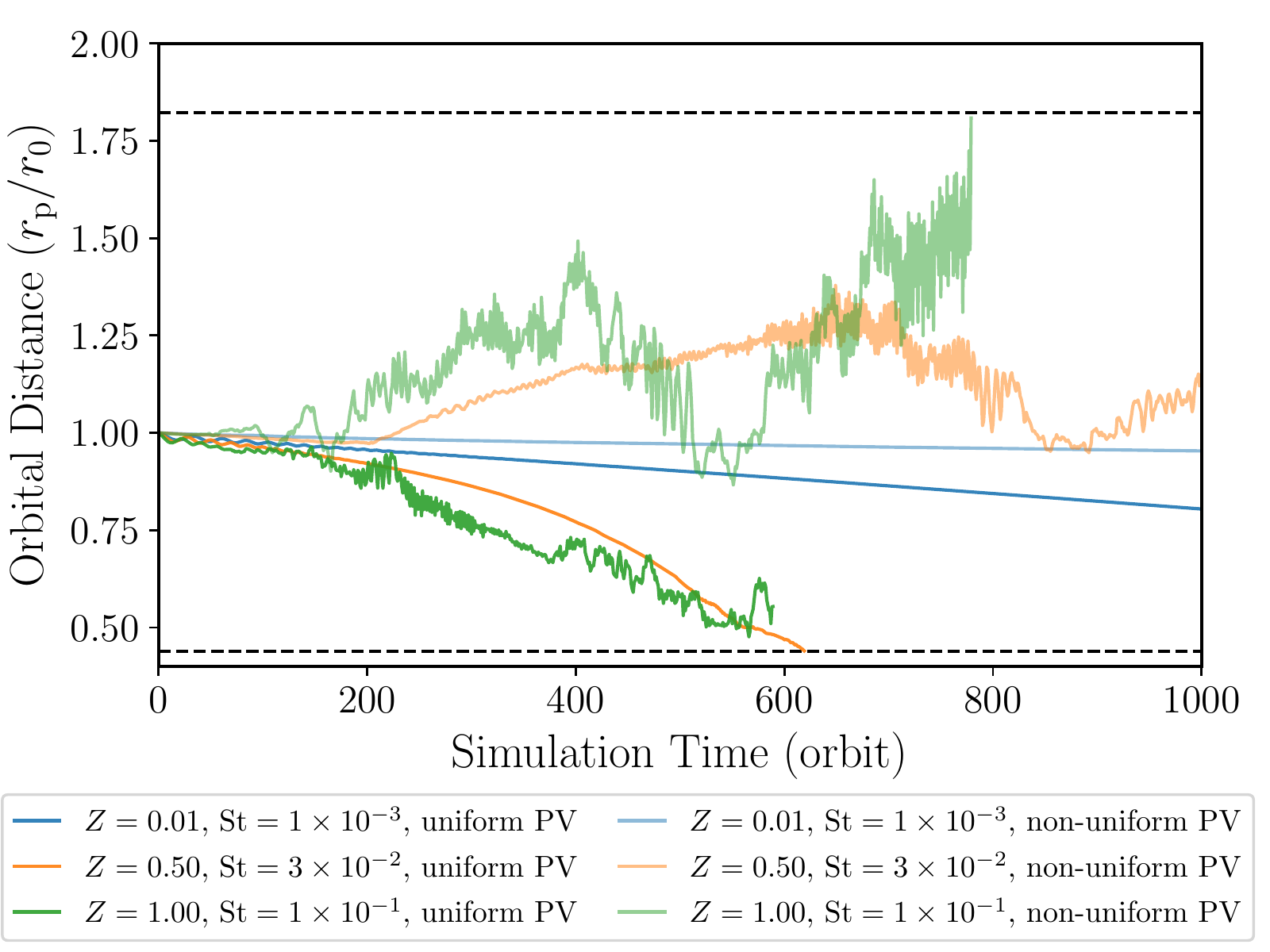}
    \caption{Orbital evolution disc models with different metallicities $Z$ and Stokes numbers $\st$. 
    Dark coloured lines correspond to a constant-PV disc and light coloured lines correspond to the fiducial model wherein the PV decreases with radius. The dashed lines mark the boundary of damping zone. Curves are terminated once the planet approach the damping zones, after which the simulation cannot be trusted. 
    \label{fig:PV_rp}}
\end{figure}

\begin{figure*}
	\includegraphics[width=2\columnwidth]{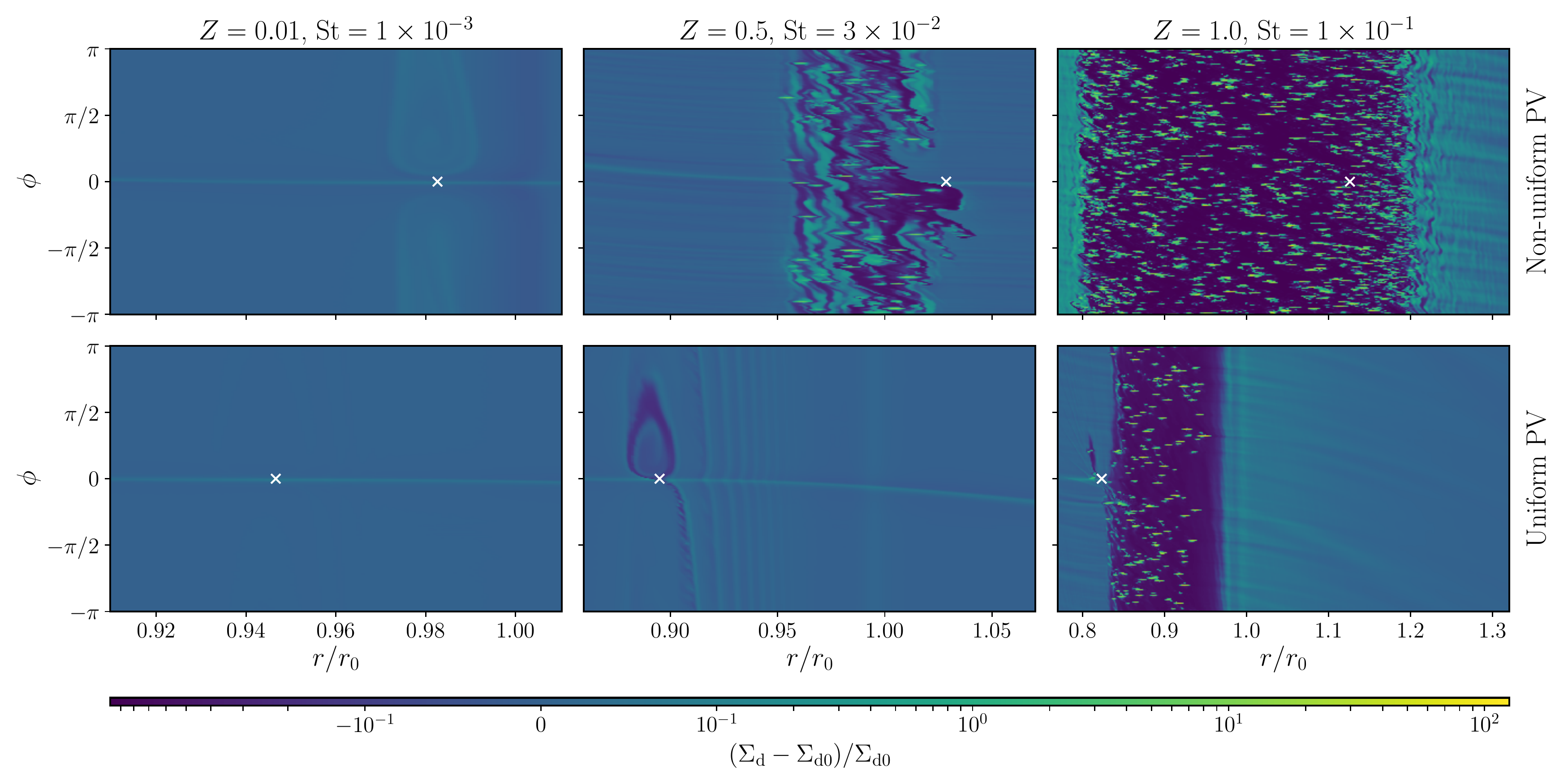}
    \caption{Relative dust surface density perturbation at 250 orbits for the disc with $Z = 0.01$ and $\st = 10^{-3}$ (left), $Z = 0.5$ and $\st = 0.03$ (middle), and $Z = 1$ and $\st = 0.1$ (right).
    The top and bottom panels show the disc model with non-uniform and uniform PV, respectively.} 
    \label{fig:PV_dens}
\end{figure*}

In Fig.~\ref{fig:PV_rp} we show the orbital evolution for three disc models with constant PV and compare them to our fiducial discs. Corresponding snapshots of normalized dust surface density at $250$ orbits are shown in Fig.~\ref{fig:PV_dens}. For $\st = 10^{-3}$, migration is steady in both cases. However, with uniform PV we find the co-orbital dust ring does not develop and the planet indeed migrates inward at a faster rate than that in the non-uniform PV disc due to the vanishing dynamical corotation torque in the former case.   

For $\st = 3 \times 10^{-2}$, which leads to stochastic migration in the non-uniform PV disc, we find that vortex formation is quenched in the constant-PV disc and the planet experiences inward, smooth runaway migration \citep{Masset2003}, which is reflected by the increasing migration rate with time. 

However, for $\st = 10^{-1}$ we find vortices still form at the outer separatrix in the uniform-PV disc, whereas vortex formation at the inner separatrix is significantly suppressed. The dust gap still forms in the vicinity of planet and evolves with time, but contains fewer dust vortices than the non-uniform PV disc. This leads to an overall inwards migration in the uniform PV disc, unlike in the non-uniform PV disc where the planet is scattered both inwards and outwards and migration is stochastic.

\subsection{Effect of Planet Mass}

We briefly consider larger planet masses with $q=3\times10^{-5}$. This corresponds to $M_\text{p} = 10M_\oplus$ around a solar-mass star, which is of interest as this is the expected minimum core mass for giant planet formation \citep{helled14}.

\begin{figure}
	\includegraphics[width=\columnwidth]{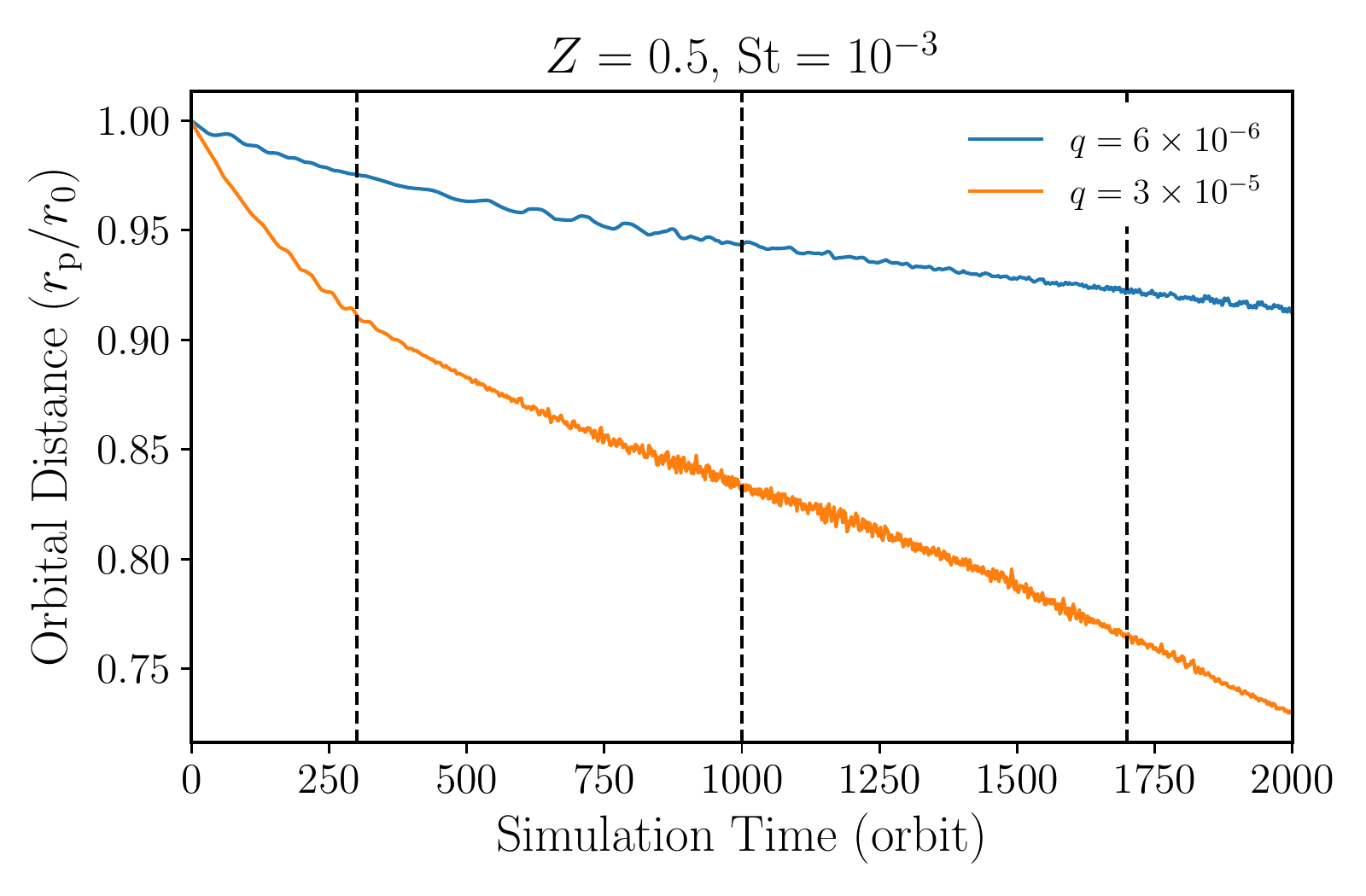}
    \caption{Orbital evolution for planet masses of $2 M_\oplus$ (blue) and $10 M_\oplus$ (orange). The disc model has $Z = 0.5$ and $\st = 10^{-3}$. The dashed lines mark the simulation times of $300$, $1000$, and $1700$ orbits
    for the panels of surface density shown in Fig.~\ref{fig:Mp_dens}.}
    \label{fig:Mp_rp}
\end{figure}

\begin{figure*}
	\includegraphics[width=2\columnwidth]{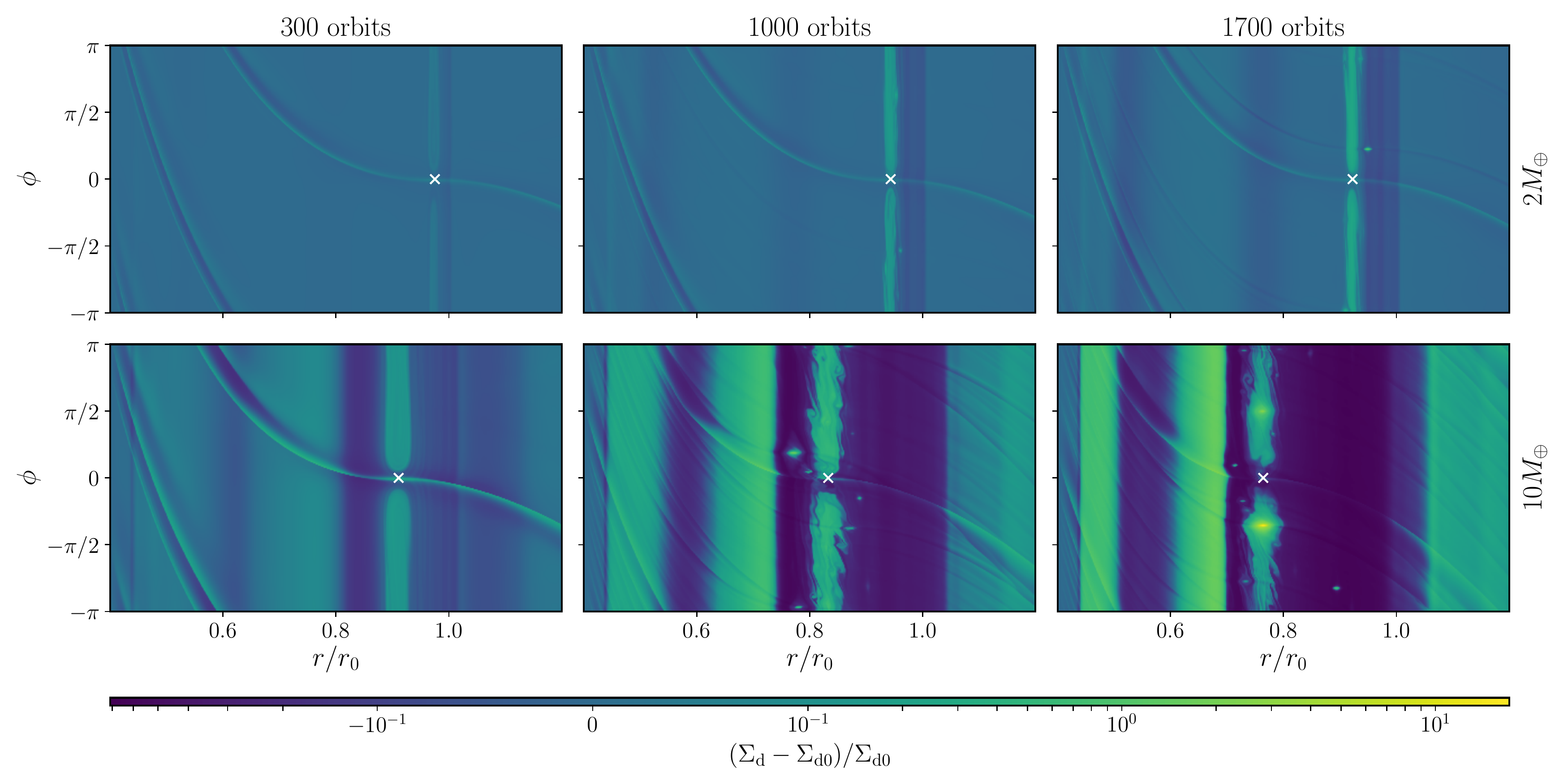}
    \caption{Relative dust surface density perturbation at $300$ orbits (left), 
    $1000$ orbits (middle), and $1700$ orbits (right)
    for a $2 M_\oplus$ (top) and $10M_\oplus$ (bottom) migrating planet in the disc with $Z = 0.5$ and $\st = 10^{-3}$.
    The planet position is marked by the white cross.}
    \label{fig:Mp_dens}
\end{figure*}

In Fig.~\ref{fig:Mp_rp} we plot the orbital evolution of a $10M_\oplus$ planet in comparison with our fiducial planet mass of $2 M_\oplus$. Here, the dust disc parameters are $Z = 0.5$ and $\st = 10^{-3}$. Evidently, the $10M_\oplus$ planet migrates inwards more rapidly with a varying migration rate. Corresponding snapshots of the dust surface density at $300$, $1000$, and $1700$ orbits are shown in  Fig.~\ref{fig:Mp_dens}. 

The $10M_\oplus$ planet opens up a double dust gap \citep{Dong2017}, which is unable to follow the migrating planet since the disc is inviscid. The planet thus migrates relative to the double gap and carries the co-orbital dust ring with it. As the dust ring pushes into the inner gap,  
the sharp PV gradient at the ring/gap edge triggers the RWI, forming dust vortices in the co-orbital region and the inner dust gap (see the middle panel of Fig. \ref{fig:Mp_dens}). The planet's migration then slows down (Fig. \ref{fig:Mp_rp}). 

As the planet migrates inward, the inner gap narrows, whereas the outer gap widens \citep{Meru2019, Perez2019, Weber2019, Kanagawa2020}. This causes the vortices in the inner gap to dissipate. In addition, vortices formed in the co-orbital region merge into two large vortices, where the vortex at the rear is larger. As a result, the migrate rate is slightly increased after $1500$ orbits. 

\begin{figure}
	\includegraphics[width=\columnwidth]{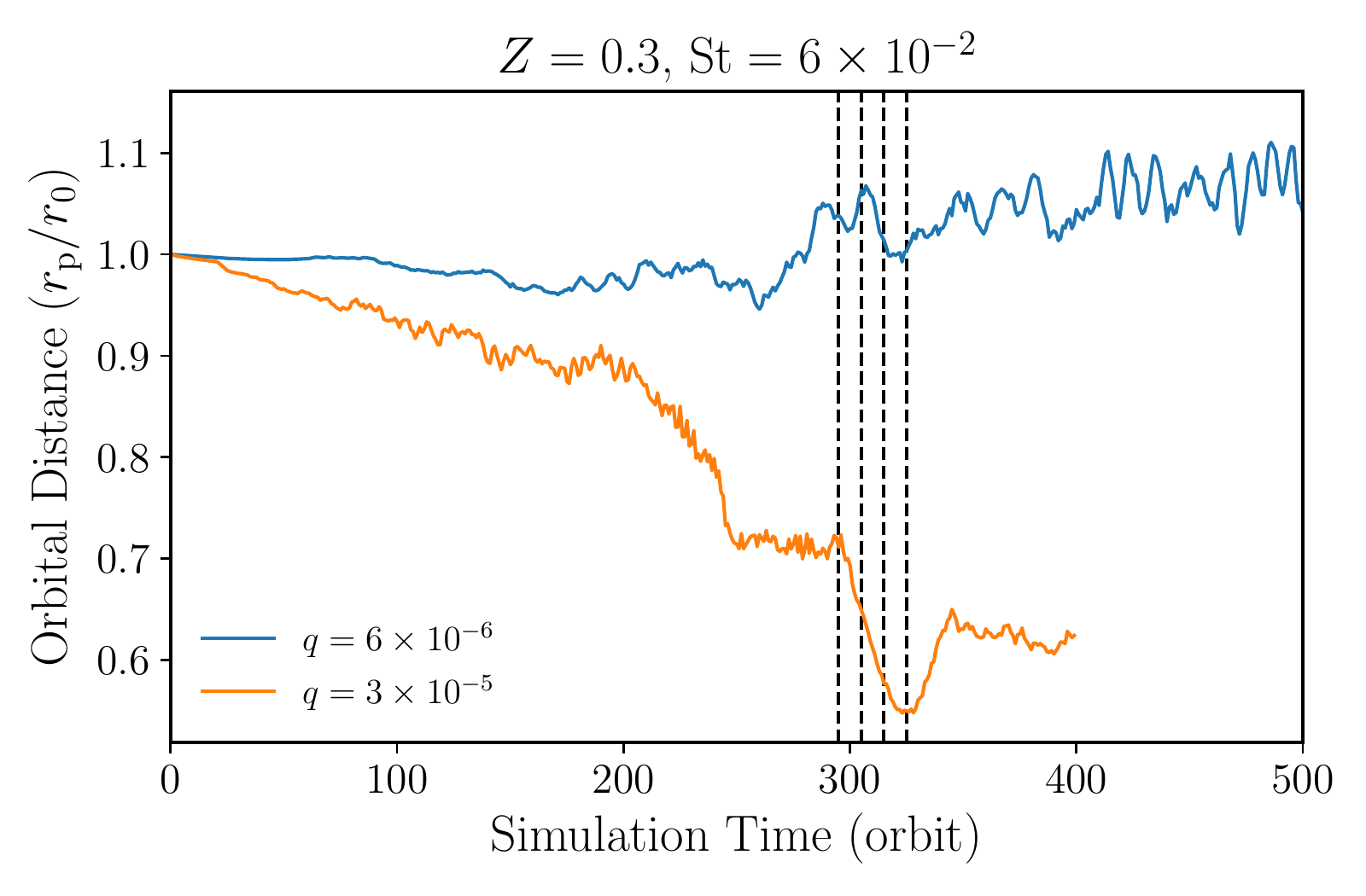}
    \caption{Same as Fig.~\ref{fig:Mp_rp}, but for the vortex-dominated regime with $Z = 0.3$ and $\st = 6 \times 10^{-2}$. For $M_p=10 M_\oplus$ (orange), the inner edge of dust gap reaches the damping zone at $400$ orbits, we thus terminate the curve afterwards. The dashed lines mark the simulation times of $295$, $305$, $315$, and $325$ orbits for the panels of dust surface density and PV shown in  Fig.~\ref{fig:Mp_dens_stochastic}.}
    \label{fig:Mp_rp_stochastic}
\end{figure}

\begin{figure*}
	\includegraphics[width=2\columnwidth]{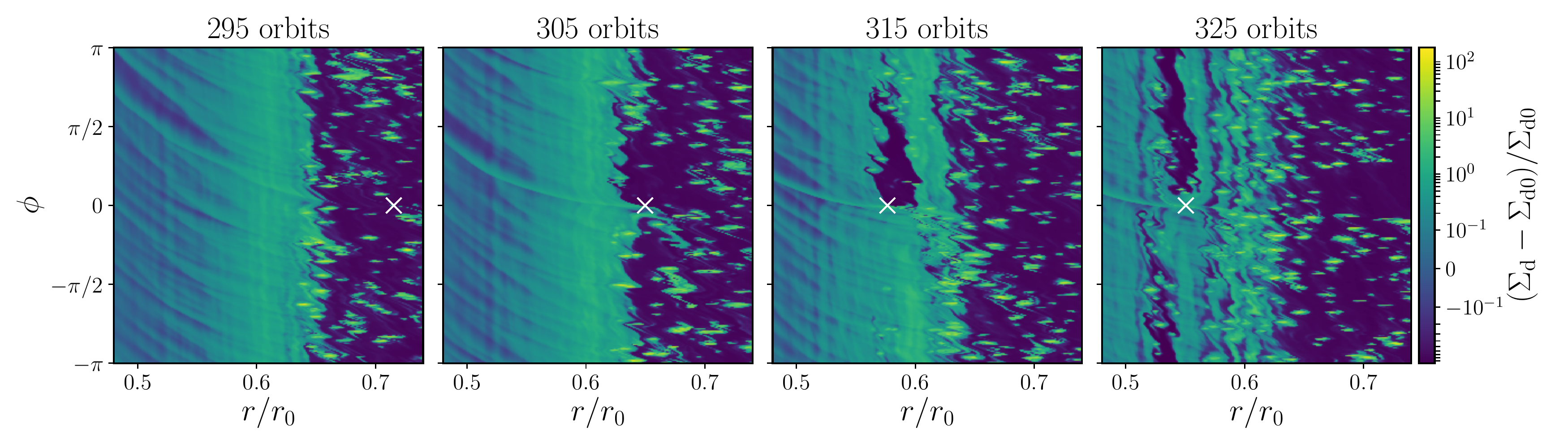}
    \caption{Relative dust surface density perturbations during an episode of rapid inwards migration of a $10 M_\oplus$ planet in the disc with $Z = 0.3$ and $\st = 6 \times 10^{-2}$.}
    \label{fig:Mp_dens_stochastic}
\end{figure*}

We also consider a disc model with $Z = 0.3$ and $\st = 6 \times 10^{-2}$, in which a $2 M_\oplus$ planet undergoes stochastic migration. Fig.~\ref{fig:Mp_rp_stochastic} shows the orbital evolution for both $2 M_\oplus$ and $10 M_\oplus$ planets. Co-orbital vortices are induced in both cases, but the $10 M_\oplus$ planet is less influenced by vortex-planet interactions, possibly due to the smaller vortex-to-planet mass ratios and the larger dust gap, which leads to a lower vortex number density and thus reduces the encounter rate with the planet. We find the planet migrates inwards overall. 
Moreover, we observe the $10M_\oplus$ planet undergoes rapid inward `jumps' at  240 and 295 orbits. In Fig.~\ref{fig:Mp_dens_stochastic} we plot the snapshot of dust surface densities before, during, and after the rapid inwards migration at 295 orbits. The planet scatters materials away from the front side and pulls materials into the rear side, leading to strong surface density asymmetries near the inner gap edge that accelerates the migration rate. This picture is similar to that found by \citet{mcnally19} for planets more massive than the feedback mass (see \S\ref{fid_model}) in inviscid, pure gas discs. However, in our dusty disc the $10 M_\oplus$ planet is below the feedback mass ($M_\text{p} \sim 0.5 M_\text{F}$) and is in the regime where dynamical corotation torques are expected to dominate.

\section{Discussion}\label{discussions}

\subsection{Observational Implications}
Our simulations suggest that low-mass planets embedded in inviscid discs can open either multiple shallow dust gaps or a single, wide, and deep dust gap filled with small-scale dusty vortices, depending on the metallicity and Stokes number (see Fig.~\ref{fig:Various_St_Z_dens}). These are expected to have distinct observational signatures.

For discs with $Z \le 0.01$ or $\st \le 10^{-3}$, we observe 
a double gap at $\sim 3-4 \widetilde{H}$ either side from the planet and two  gaps located at $r \le 0.6 r_0$ ($\sim8 \widetilde{H}$), which is consistent  with \citet{Dong2018}. These 4 gaps become more visible in discs with larger Stokes number and develop for both migrating and non-migrating planets. However, for migrating planets we also find a significant gas surface density bump is generated in front of the planet, which leads to  
the formation of a co-orbital dust ring and a `detached' dust gap located just outside the planet (see Fig.~\ref{fig:St1e-3_Z0.01_densave_nonmig_mig}), which can have comparable depths to the 4 gaps discussed above. Moreover, we find the outer edge of detached gap does not significantly evolve with the migrating planet, so that it may be used to estimate the initial position of planet.

When $Z \ge 0.03$ and $\st \ge 3 \times 10^{-2}$, a single, deep and wide dust gap is formed, which is filled with dust vortices with radial length-scales ranging from $0.01 \widetilde{H}$ to $1 \widetilde{H}$. The metallicity within vortices can increase up to $O(10^2)$ as they collect more dust over time. These dust vortices induce spiral structures in both the gas and dust, forming two feathered-rings at the gap edges. We find that the migrating planet resides within the dust gap most of the time, but its precise location is difficult to pinpoint due to the lack of a co-orbital dust ring.

Our simulations suggest observations of multiple narrow dust rings and gaps 
may be explained by a low-mass planet migrating in an inviscid disc with low $Z$ and small $\st$. In this case the planet would be located at a dust {\it ring}, rather than inside a dust gap.  
On the other hand, a single wide dust gap can be induced a single low-mass planet in discs with high $Z$ and larger $\st$, as opposed to gap-opening by multiple giant planets \citep{Zhu2011}.

\subsection{Implications for Planet Migration}

We find that planet migration becomes chaotic in discs with $Z\gtrsim 0.3$ or $\st\gtrsim 0.03$ due to the formation of numerous co-orbital vortices that scatter the planet, with no clear trend for inwards or outwards migration over our simulation time-scales. This is in contrast to inwards migration observed in discs with smaller $\st$ and $Z$. This suggests that in realistic disc where both $Z$ and $\st$ can evolve, that orbital migration of low-mass planets may transition from smooth and inwards to stochastic.

Consider, for example, a dusty disc with $Z = 0.01$ and $\st = 10^{-3}$ initially, for which our simulations show that the planet migrates inward steadily with a dust ring formed in the co-orbital region. The metallicity in the ring increases with time as particles accumulate. Particles are also expected to grow in size \citep{Drazkowska2019,Li2020}. Our results suggest that when $Z \gtrsim 0.1$ and $\st \gtrsim 10^{-2}$ in the dust ring, vortices would form in the vicinity of planet and slow down its migration. For $Z\gtrsim 0.3$ and $\st\gtrsim 0.03$, the  continuous development of vortex instabilities at the edges of the dust gap widens it. The planet then undergoes stochastic migration due to vortex-planet scattering, but is constrained to the dust gap.  

The long-term outcome from the interaction between the migrating planet and co-orbital vortices is difficult to predict. One possibility is that planetesimals may form within the dust vortices as they have high metallicities \citep{Youdin2005}, leading to the formation of planetary embryos or protoplanets. Together with the original planet, this may be one way to form multi-planet systems. However, the interaction between the planets and remaining small-scale dust vortices could disrupt resonant chains within the multi-planet system \citep{McNally2019b}.

\subsection{Caveats}

It is important to keep in mind several simplifications made in our disc models, namely (1) the thin-disc approximation; (2) flat pressure profiles and simplified thermodynamics; (3) non-self-gravitating discs; and (4) the fluid treatment of dust particles with constant Stokes number. We discuss these caveats below. 

In realistic, three-dimensional (3D) discs, dust settles vertically \citep{dubruelle95} so that the local dust-to-gas mass density ratio declines with height away from the disc midplane. Thus, the degree of particle feedback onto the gas also falls with height, but this decrease is neglected in 2D models that only consider surface densities. Our 2D models likely over-estimate the effect of particle feedback compared to 3D discs with the same $Z$. 

Alternatively, 2D models may be regarded as representing the midplane of a 3D disc. 
If the dust layer is thin, as would be the case for the weakly turbulent discs considered in this work, the total mass contained within the dust layer would also be limited. This means that, while phenomena such as vortex formation should still occur near the midplane, the planet's migration would still be dictated by the vortex-free gas layers above and below. In this interpretation, we expect low-mass planets to undergo standard inwards type I migration \citep{tanaka02} with minor perturbations from the midplane dusty vortices.

The above issues are best resolved via direct 3D simulations of dusty disc-planet interaction. In this case, one would also need to include particle stirring by simulating gas turbulence \citep[e.g.][]{flock17,flock20,lin19,schafer20} to maintain a finite dust layer thickness. The cost of such 3D simulations likely prohibits a wide parameter survey, but our 2D results suggest the regime of interest, as far as planet migration is concerned, will be large  metallicity and Stokes number (i.e. $Z\gtrsim0.3$, $\st\gtrsim 0.03$).

We employed a special disc profile in which there is no radial drift of dust particles initially. In more typical discs with a negative pressure gradient, inwardly drifting particles can cross the planet's co-orbital region if the particle drift velocity exceeds the planet's migration rate. This radial mass flux may exert additional torques on the planet \citep{mcnally17}. It can also enhance the metallicity in the co-orbital region, which can become unstable, implying that planet  migration can evolve from steady to stochastic over time. Future work should explore more general pressure profiles. 

Our isothermal discs correspond to the limit of instant cooling. In discs with cooling timescales comparable with the orbital period, a wide gap is formed around the planet, and disc-on-planet torque is modestly reduced \citep{Miranda19,Miranda20}. This may reduce the metallicity in the co-orbital region and delay vortex formation.
However, once the planet enters the vortex-dominated regime, we expect disc thermodynamics to have limited effect on orbital migration, i.e. it will still be stochastic. Nevertheless, this should be checked with simulations that include the energy equation and explicit cooling.

In our simulations the planet feels the disc potential, but the disc itself is non-self-gravitating. The mismatch can lead to an over-estimate of the differential Lindblad torque for freely-migrating planets \citep{baruteau08}. We confirmed this effect with additional simulations that account for this discrepancy\footnote{By removing the azimuthal component of the disc surface density prior to force evaluations, see https://fargo3d.bitbucket.io/nbody.html.}. However, this did not significantly affect the overall migration behaviour, especially when either dynamical corotation torques or vortex-planet interactions dominate. On the other hand, a fully self-gravitating disc may provide an effective viscosity that suppresses vortex formation and prevents stochastic migration \citep{Pierens2019} -- an effect that will depend on the disc mass and should be examined further.

Finally, the fluid treatment of dust with constant Stokes number can be improved. The Stokes number should be inversely proportional to the gas surface density \citep{weiden77}, so its value should also evolve with the evolving gas disc. Specifically, at gap edges where the gas surface density is enhanced, $\st$ should decrease, which may suppress vortex formation and reduce the tendency for stochastic migration. In our simulations the gas surface density varies by less than a factor of two from its initial values, so this is not expected to have a significant impact, but may become important for more massive planets that carve deep gaps and induce strong pressure bumps. 

The fluid approximation also does not allow for crossing particle trajectories or particle-particle interactions, which can be expected in the high-metallicity dust vortices in our simulations. By comparison to \citet{Yang2020}, who used Lagrangian particles, we find that vortex formation appears to be less efficient in the fluid approach, especially in discs with small Stokes number. Thus, transition to vortex-dominated, stochastic migration may occur at lower values of $Z$ and $\st$ than that identified in our fluid-based approach.

\section{Summary and conclusions}\label{summary}

In this paper, we study planet migration in inviscid, dusty protoplanetary discs using a series of 2D simulations. We treat the dust as a separate, pressureless fluid and include the dust feedback onto the gas. 
We mostly consider low-mass planets with $M_\text{p} = 2 M_\oplus$ around solar-mass star and systematically study its migration a function of 
the disc's initial dust-to-gas mass ratio $Z$ (or metallicity, with $0.01 \le Z \le 1$) and particle Stokes numbers $\st$ (with $10^{-3} \le \st \le 10^{-1}$).
We also briefly explore the effect of viscosity, resolution, surface  density profile, and planet mass. 

For discs with low metallicity ($Z \lesssim 0.1$) and/or small Stokes number ($\st \lesssim 10^{-2}$), the planet migrates inward 
but slows down due to dynamical corotation torques from the gas disc \citep{Paardekooper2014}. We observe that a co-orbital dust ring and a detached dust gap are formed in the vicinity of the migrating planet. These co-orbital features are much less prominent without planet migration. At moderate metallicities or Stokes numbers, e.g. $Z \sim 0.15$ for $\st = 10^{-3}$ and $Z \sim 0.03$ for $\st \ge 10^{-2}$, we find small-scale fluctuations in the planet's orbital evolution due to periodic asymmetries in the librating dust flow near the planet. However, these do not affect the overall tendency for inwards migration.

However, we find that planet migration becomes chaotic in discs with high metallicity ($Z \gtrsim 0.3$) and large Stokes number ($\st \gtrsim 0.03$). In these cases, small-scale dust vortices first develop in the planet's co-orbital region, but later also develop at the edges of the dust gap initially carved by the planet, which further widen the dust gap. These vortices continuously scatter the planet, which can halt or even reverse its initial inwards migration, but the long term outcome of stochastic migration is unclear. Nevertheless, we find the planet is constrained to reside within the dust gap due to a `repulsive' dynamical corotation dust torque 
that pushes the planet back into the dust gap whenever it is scattered outwards (inwards) from the outer (inner) gap edge.

The stochastic migration regime we identify relies on efficient vortex formation in the vicinity of planet, which also depends on the resolution, viscosity, and the initial vortensity or potential vorticity (PV) profile. We find a resolution of $\gtrsim (90, 40)$ cells per scale-height in the radial and azimuthal directions, respectively; an alpha viscosity $\alpha \lesssim 3 \times 10^{-5}$; and  non-uniform PV profiles are needed to enter this regime. These requirements stem from the fact that the initial vortex instabilities are driven by localised, steep, co-orbital PV gradients that must be resolved and can only be generated if the background PV is non-uniform. We also find that super-Earths ($M_\text{p} = 10 M_\oplus$) trigger the dusty vortex instabilities more easily, but the vortices do not strongly affect the planet's overall inwards migration, though it can be considerably non-smooth. 

Our simulations show that low-mass planet migration in weakly turbulent, dust or pebble-rich discs are strongly affected by vortices and are likely stochastic. Thus, it may not be appropriate to apply type-I migration torque formula \citep[e.g.][]{Paardekooper2010} under  dust-rich conditions. Instead, a statistical approach with extended integration times may be needed to ultimately assess the impact of dust-induced planet migration on the formation of planetary systems.

\section*{Acknowledgements}
We thank Colin McNally for a timely and constructive report. We also thank Arnaud Pierens for comments and suggestions on an early draft of this paper. This work is supported by the Ministry of Science and Technology of Taiwan through grants 107-2112-M-001-043-MY3, 107-2112-M-007-032-MY3, and 108-2811-M-007-562. Numerical simulations were performed on the CICA cluster at the National Tsing Hua University, as well as the Taiwan Computing Cloud at the National Center for High-performance Computing (NCHC). The CICA cluster was funded by the Ministry of Education of Taiwan, the Ministry of Science and Technology of Taiwan, and National Tsing Hua University. We are grateful to the NCHC for computing time,  facilities, and support.

\section*{Data Availability}

All data generated during this study are included in this published article.




\bibliographystyle{mnras}
\bibliography{mnras} 



\appendix

\section{Code comparison}\label{pluto_compare}

In our previous work \citep{Chen2018} we treated the dusty-gas disc as a single fluid system by considering  tightly-coupled dust and applying the terminal velocity approximation \citep{Lin2017,lovascio19}. We here compare dusty disc-on-planet torques obtained from this approach and implemented in the \textsc{pluto} code \citep{mignone07,mignone12} by \citeauthor{Chen2018} to the two-fluid approach adopted 
in the \textsc{fargo3d} code and used in the main text. 

We consider a non-migrating $2 M_\oplus$ planet in our fiducial disc model (\S\ref{model}) with metallicity $Z = 0.5$ and Stokes number $\st = 10^{-3}$. The \textsc{pluto} setup assumes a fixed particle size such that $\st\propto \Sigma_\text{g}^{-1}$ instead of a strictly constant $\st$, but this is inconsequential because gas surface densities perturbations remain small. We use the same simulation domain and reoslution as \citeauthor{Chen2018}:
$(r, \phi) \in [0.6, 1.4] \times [0, 2 \pi]$ with $(N_r, N_\phi) = (720, 2672)$. 

\begin{figure}
	\includegraphics[width=\columnwidth]{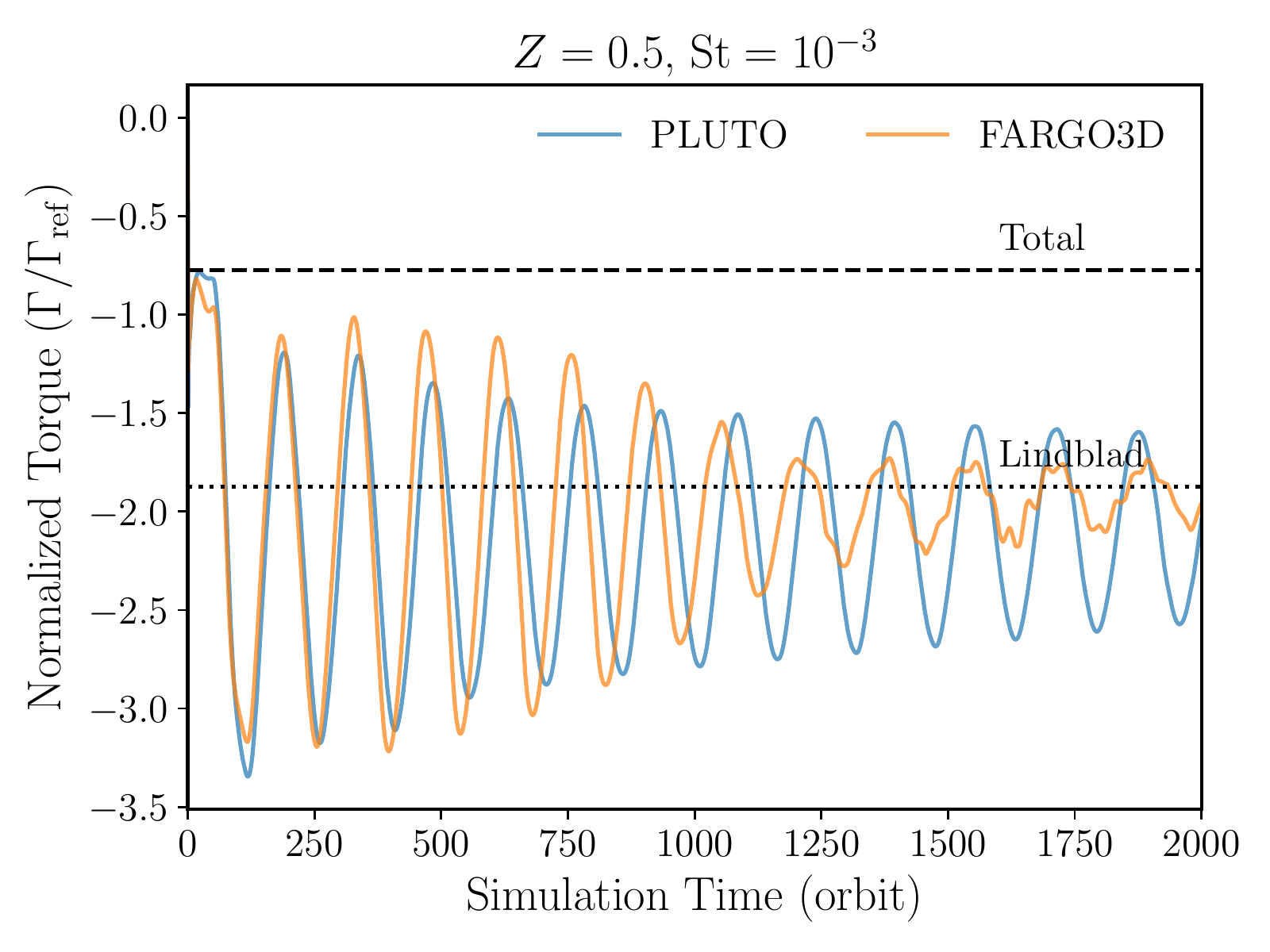}
    \caption{Normalized disc-on-planet torque for a non-migrating $2 M_\oplus$ planet 
    in the fiducial disc with $Z = 0.5$ and $\st = 10^{-3}$. 
    The blue line is the resulting torque from the single-fluid approach using \textsc{pluto},
    and the orange line is from the multi-fluid approach using \textsc{fargo3d}.
	The dotted and dashed lines are the semi-analytical values 
    of Lindblad torque and total disc-on-planet torque, respectively.}
    \label{fig:code_comparison}
\end{figure}

In Fig.~\ref{fig:code_comparison} we show the disc-on-planet torque normalized by the reference torque. The resulting torques are consistent at the first $1000$ orbits in both approaches. The large-scale torque oscillations result from 
periodic co-orbital density asymmetries that librates about the planet  \citep{Chen2018}.

However, in the \textsc{fargo3d} run the co-orbital flow becomes turbulent after $1000$ orbits, which causes the co-orbital asymmetry to dissipate and the torque converges to the Lindblad values. On the other hand, the \textsc{pluto} run remains stable. This suggests the \textsc{pluto} setup is suffers from larger numerical diffusion that prevents the development of vortex instabilities at the chosen $Z$ and $\st$.

\section{Dynamical Corotation Torques in dusty discs} \label{append:torque_constMigRate}

We conduct a series of customised simulations to study the effect of migration rate on the dynamical corotation torque from dust disc. We modify the \textsc{fargo3d} code to force the planet to migrate inwards at a constant rate by applying an additional specific force in the azimuthal direction: 
\begin{equation}
F_{\phi, \text{forced}} = \frac{\Omega_\text{K} (r_\text{p})}{2} \frac{d r_\text{p}}{dt}.
\end{equation}
The migration rate is parameterized in form of
\begin{equation} \label{eq:MigRate}
\frac{d r_\text{p}}{dt} = -\mathcal{C} \frac{2 x_s}{\tau_\text{lib}} 
                        = -\mathcal{C} \frac{3 \Omega_\text{p} x_s^2}{4 \pi r_\text{p}}, 
\end{equation}
where $x_s$ is the half-width of horseshoe region given by equation~(\ref{paar_xs}) and $\tau_\text{lib}$ is the libration time-scale. 
The factor $C$ distinguishes slow $(C < 1)$ and fast migrations $(C > 1)$ by comparing the time the planet takes to migrate across the horseshoe region to the libration time-scale.
For simplicity we set the right-hand-side of equation~(\ref{eq:MigRate}) to its value at the planet's initial radius $r_0$. We consider the fiducial disc model with $Z = 0.3$ and $\st = 6 \times 10^{-2}$, and explore $C = 0$ (no migration), $2$, $4$, and $8$.

\begin{figure*}
	\includegraphics[width=2\columnwidth]{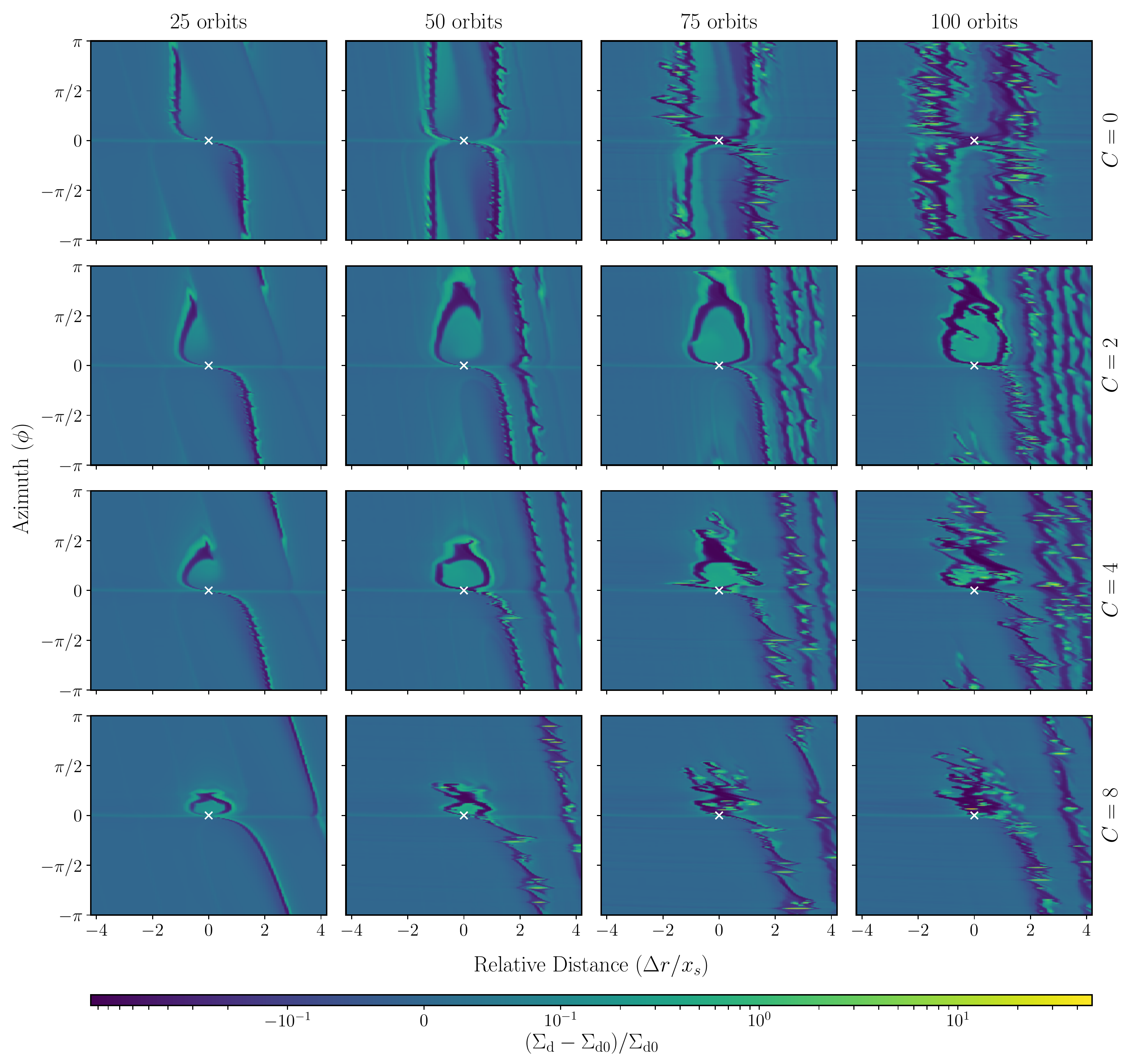}
    \caption{Snapshot of dust density at different simulation times for a forced migrating planet in a disc with $Z = 0.3$ and $\st = 6 \times 10^{-2}$.
    The panels are organised such that the simulation time increases from left to right,
    and the migrate rate increases from top to bottom.}
    \label{fig:ForcedMig_dens}
\end{figure*}

\begin{figure*}
	\includegraphics[width=2\columnwidth]{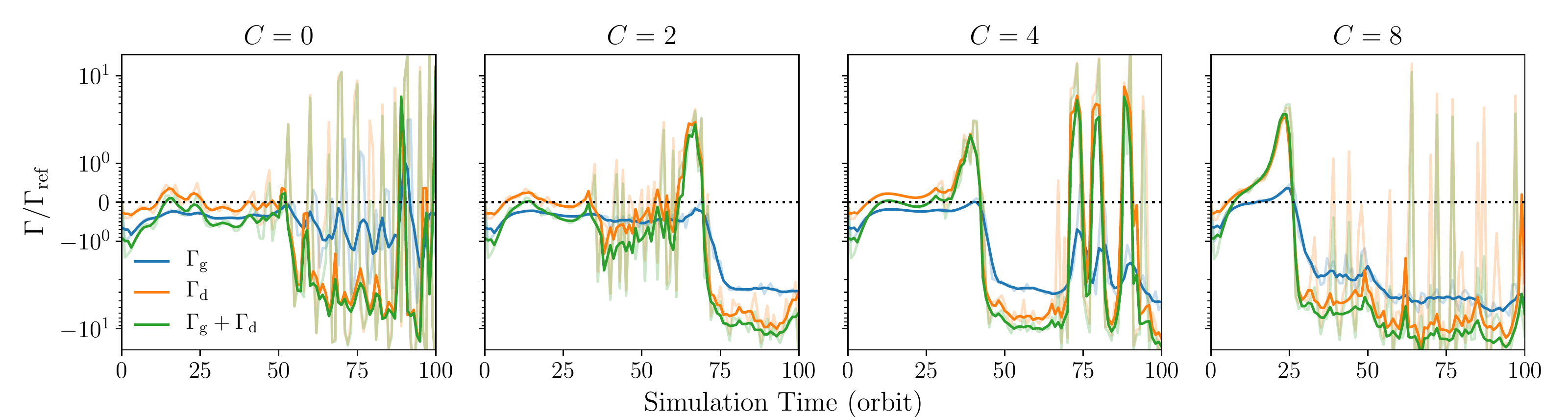}
    \caption{Evolution of disc-on-planet torque.
    Light coloured lines show the instantaneous torque, 
    and dark coloured lines show the 5-orbit running average.}
    \label{fig:ForcedMig_torq}
\end{figure*}

Fig.~\ref{fig:ForcedMig_dens} shows snapshots of the dust surface density in the vicinity of planet and Fig.~\ref{fig:ForcedMig_torq} shows the evolution of disc-on-planet torque. As pointed out in the main text, in simulations with $C > 0$, a libration island is formed at the front side of the inwardly migrating planet, with an overdense dust flow located outside the island. The torque exerted by the overdense flow is positive as it approaches to the planet from the outer upstream separatrix, but becomes negative once it moves past and being scattered by the planet.

The above picture is similar to that found by \citet{llambay18} who considered a non-migrating planet but with an radially inwards pebble flux, which is equivalent to an outwardly migrating planet without a radial pebble flux. In that case the drifting pebble flow generates a libration island at the rear side and exerts positive torque. 

We next focus on the disc-on-planet torque before the overdense flow is scattered, which occurs at about $70$ ($C = 2$), $40$ ($C = 4$), and $20$ orbits ($C = 8$). For $C \ge 4$, the dust torque can overwhelm the negative gas torque, resulting in the positive total disc-on-planet torque. At higher migration rates, the positive total torque is much stronger, which can slow down the migration on short  time-scales. In the more realistic case of a freely migrating planet,  the overdense flow at the front side would move into the libration island and perform horseshoe turns as the planet is slowed down, which exerts a positive torque that can further halt the migration or even reverse its direction.

\bsp	
\label{lastpage}
\end{document}